\documentclass[prb,a4,
              twocolumn,
              showpacs,amsmath,amssymb]{revtex4-1}
\usepackage{epsfig}
\usepackage{graphicx}
\usepackage{subfigure}
\usepackage{color}
\usepackage{soul}
\usepackage{float}
\usepackage{epstopdf}
\usepackage{amsmath}
\usepackage{amssymb}
\usepackage{ifthen}
\usepackage{alltt}
\usepackage{fancyhdr}
\usepackage{verbatim}
\usepackage{moreverb}
\usepackage{rotating}
\usepackage{colortbl}
\usepackage{amssymb}
\usepackage{multirow}
\usepackage{bigstrut}
\usepackage{wrapfig}

\def\e{\begin{equation}}
\def\f{\end{equation}}
\def\_#1{{\bf #1}}

\def\.{\cdot}

\def\Re{{\rm Re\mit}}

\def\aeeoh{\widehat{\alpha}_{\rm ee}}
\def\ammoh{\widehat{\alpha}_{\rm mm}}
\def\aee{\alpha_{\rm ee}}
\def\amm{\alpha_{\rm mm}}
\def\be{\beta_{\rm e}}

\def\=#1{\overline{\overline #1}}

\begin{document}

\title{Broadband reflectionless metasheets:\\
 Frequency-selective transmission and perfect absorption}
\author{V.~S.~Asadchy$^{1,2}$, I.~A.~Faniayeu$^{2,3}$, Y.~Ra'di$^{1}$, S.~A.~Khakhomov$^{2}$, I.~V.~Semchenko$^{2}$, and S.~A.~Tretyakov$^{1}$}
\affiliation{$^1$Department of Radio Science and Engineering, Aalto University, P.~O.~Box~13000, FI-00076 Aalto, Finland\\
$^2$Department of General Physics, Francisk Skorina Gomel State University, 246019 Gomel, Belarus\\
$^3$Research Institute of Electronics, Shizuoka University 3-5-1 Johoku, Naka-ku, Hamamatsu 432-8011, Japan}


\begin{abstract}
Energy of propagating electromagnetic waves can be fully absorbed in a thin lossy layer, but only in a narrow frequency band, as follows from the causality principle. On the other hand, it appears that there are no fundamental limitations on broadband matching of thin absorbing layers. However, known thin absorbers produce significant reflections outside of the resonant absorption band. In this paper we explore possibilities to realize a thin absorbing layer which produces no reflected waves in a very wide frequency range, while the transmission coefficient has a narrow peak of full absorption.   Here we show, both theoretically and experimentally, that a wide-band-matched thin resonant absorber, invisible in reflection, can be realized if one and the same resonant mode of the absorbing array unit cells is utilized to create both electric and magnetic responses. We test this concept using  chiral particles in each unit cells, arranged in a periodic planar racemic array, utilizing chirality coupling in each unit cell but compensating the field coupling at the macroscopic level. We prove that the concept and the proposed realization approach also can be used to create non-reflecting layers for full control of transmitted fields. Our results can have a broad range of potential  applications over the entire electromagnetic spectrum including, for example, perfect ultra-compact wave filters and selective multi-frequency sensors.

\end{abstract}

\maketitle

\section{Introduction}
Total absorption of electromagnetic radiation requires elimination of all wave propagation channels: reflection, transmission, and scattering. Here we study electrically thin absorbing sheets, which are of prime importance for many applications, for example, wave filtering, radar cross section reduction, energy harvesting, sensing, thermal emission control. It is known that incident electromagnetic energy can be nearly fully absorbed in thin layers \cite{perfect,infrared,soukoulis,munk2}, but only in a narrow frequency band. The absorption bandwidth of any passive layer obeys a fundamental limitation, as follows from the causality principle \cite{Rozanov,bounds_tr,bounds_tr2}. In fact, the maximal bandwidth of absorption is proportional to the layer thickness \cite{Rozanov}. This limitation can be overcome only using active (pumped by some external power source) structures. 

On the other hand, apparently it has not been noticed before that there is no such fundamental limitation on the frequency range in which the \emph{reflection} from a thin resonant absorbing layer can be made  negligible. Conceptually, we do not see any reason why it would be impossible to realize a thin layer which fully absorbs the incident power in a narrow frequency band and allows the wave to freely pass through at other frequencies, thus, producing no reflections at all (within the band where the layer remains electrically thin and the inclusions forming the absorber remain electrically small). 
The existence of such a structure does not contradict known fundamental limitations. 
Indeed, the classical limitation on matching bandwidth \cite{bounds_Fano} applies only to   lossless matching networks. The limitation on the absorption bandwidth \cite{Rozanov} holds only for absorbers containing an impenetrable mirror. Thus, these restrictions are not applicable for the proposed lossy structure lacking a ground plane. The limitations on periodical arrays \cite{bounds_tr2} concern only the transmission properties.  Obviously, exploitation of the opportunity to design a wide-band matched absorber could open up a number of novel possibilities in applications, for example, in ultra-thin filters for wave trapping, selective multi-frequency bolometers and sensors. Such an all-frequencies-matched absorber would be ``invisible'' from the illuminated  sides, still acting as a band-stop filter in transmission. To the best of our knowledge, such wideband matched thin absorbers are not known. The main goal of this paper is to explore a possibility to realize such structures. 

In fact, most of the known designs of thin absorbers contain a continuous metal ground plane (a mirror) behind the absorbing layer \cite{perfect,infrared,soukoulis,munk2}. The mirror obviously produces nearly full reflection outside of the absorption band.  The use of a mirror reflector can be avoided in absorbers based on arrays of subwavelength Huygens' sources (so-called Huygens' metasurfaces) that possess the appropriate level of dissipative loss. Such Huygens' sources scatter secondary waves only in the forward direction (without reflection) which destructively interfere with the incident wave, yielding zero transmission.
Pioneering topologies of Huygens' inclusions were introduced in Refs.~[\onlinecite{kerker,samofalov,ozbay,sks}]. Subsequently, Huygens' inclusions of different topologies have been used as structural elements in sheets to control transmission wavefronts \cite{Huygens1,Huygens2,Huygens3}.
Recently, there have been proposed several topologies of absorbers based on cut wire arrays separated by a dielectric layer \cite{ref1,ref2,ref3,ref4,7padilla}. However, in all these structures the Huygens' balance between the electric and magnetic responses (which is necessary for cancellation of the reflected waves) holds only inside a narrow-frequency band for which the dimensions have been optimized. Outside of this band, reflections appear due to prevailing excitation of either electric or magnetic modes. The physical reason for this is that the different resonant modes exhibit different frequency dispersions.  The same conclusion is valid for the idea of using a resistive sheet placed in a close proximity of resonating magnetic inclusions to realize absorbing layers \cite{bilotti}. 
The analysis of these solutions reveals another problem of the known designs: The absorbers are realized as rather complicated multi-layered structures. Typically, they  comprise at least three layers (metal--dielectric--metal). 
In addition to  practical advantages, it is of primary theoretical importance to engineer an absorber which comprises only a \emph{single-layer} metasheet. This essentially allows one to build extremely thin absorbers. 

Furthermore, it is important to note that all the polarization-insensitive absorbers, which have been proposed so far, operate only for waves incident on only one side of the absorbing layer or when the two sides are illuminated by two coherent waves.
The underlying reasons of this are the presence of a ground plane (in metal-backed structures) and bianisotropic effects \cite{serdukov,absorption} (in structures without a ground plane) destroying the absorption symmetry. Moreover, this asymmetry compromises the desired invisibility of the absorbers in reflection from both sides.
In our previous work \cite{absorption} we demonstrated that total \emph{symmetric} absorption in an electrically thin layer can be achieved only if the layer is not bianisotropic (there is no electromagnetic coupling in the layer).
The only exception of achieving two-sided absorption with uncompensated bianisotropy was theoretically predicted in Ref.~[\onlinecite{absorption}], however, in that case the absorbing layer is polarization-sensitive. Subsequently, an example of such a symmetric and polarization-sensitive absorber has been reported in Ref.~[\onlinecite{CP}], providing absorption only for one circular polarization (50\% of incident power).

In this paper a possibility to create a thin resonant polarization-insensitive absorber which produces negligible reflections in an ultra-wide frequency range is explored and experimentally demonstrated. We examine the physical  requirements for full and symmetric (from either of the sides) absorption of incident waves and show that the ideal performance can be accomplished in a single-layer array of specifically designed helical inclusions. We show, both theoretically and experimentally,  that an array of these inclusions truly operates as a Huygens' surface in a very wide frequency range, exhibiting zero reflectivity even far from the absorption band. The governing idea behind this solution is excitation of both electric and magnetic surface currents using just one resonant mode of complex-shaped inclusions.  

\section{Symmetric total absorption in a single-layer array of resonators}
\begin{figure}[h]
\centering
\epsfig{file=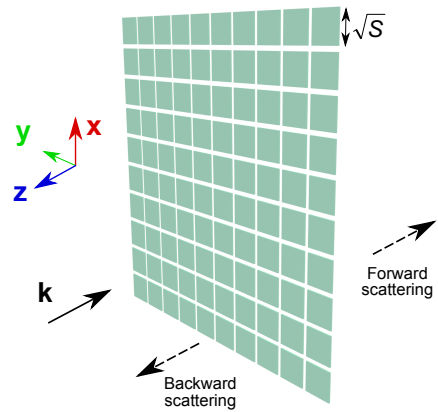, width=0.65\columnwidth}
\caption{(color online) A schematic illustration of a generic, electrically thin metasheet.}
\label{ris:fig1}
\end{figure}

Consider a uniaxial (possessing uniaxial symmetry in the plane) metasurface (or, rather, a \emph{metasheet}) formed by a single two-dimensional periodic array of identical subwavelength inclusions. The inclusions are both electrically and magnetically polarizable. The subwavelength size of the inclusions and the subwavelength period allows us to describe the electromagnetic response of the metasurface in terms of electric and magnetic dipole moments induced in the inclusions. The higher-order moments do not influence the reflection and transmission coefficients for plane-wave excitation. The array is illuminated by a normally incident plane wave propagating along the $-{\bf z}$ direction, as shown in Fig.~\ref{ris:fig1}.
Since the array period is smaller than the wavelength of the incident radiation, the dipole moments induced in the inclusions can be modeled as surface-averaged electric and magnetic current sheets which radiate secondary plane waves in the backward and forward directions. The backward scattered waves form reflection, whereas interference of the forward scattered and the incident waves defines transmission (see Fig.~\ref{ris:fig1}). As shown in Ref.~[\onlinecite{Teemu}], the electric fields of the reflected and transmitted waves from a metasurface illuminated normally by an incident plane wave are given by
\begin{equation}
\begin{array}{l}
\displaystyle
\mathbf{E}_{\rm r}=\frac{i\omega}{2S} \left(\eta_0\aeeoh-\frac{1}{\eta_0} \ammoh \right) \mathbf{E}_{\rm inc},\vspace*{.2cm}\\\displaystyle
\mathbf{E}_{\rm t}= \left[1+\frac{i\omega}{2S}\left(\eta_0\aeeoh+\frac{1}{\eta_0}\ammoh\right)\right] \mathbf{E}_{\rm inc},\\\displaystyle
\end{array}\label{eq:q1}
\end{equation}
where $\omega$ is the angular frequency, $S$ is the area of the array unit cell, $\eta_0$ is the free-space wave impedance, $\aeeoh$ and $\ammoh$ are the effective (collective) electric and magnetic polarizabilities of the unit cells. Relations~(\ref{eq:q1}) hold for incident waves impinging on the metasurface from $+\bf z$ and $- \bf z$ directions. 
It should be stressed that the symmetric absorption regime in a metasurface is possible only if there is no bianisotropic coupling in the structure \cite{absorption}. To the best of our knowledge, all known thin single-layer absorbers demonstrate asymmetric response caused by bianisotropy of the structure. 
In this work, we focus on an absorber design that ideally operates for incident waves impinging on either or both of its sides and show how the bianisotropy can be suppressed. 

Requiring $\mathbf{E}_{\rm r}=0$ and $\mathbf{E}_{\rm t}=0$ in Eqs.~(\ref{eq:q1}), one can find the  conditions of symmetric total absorption in a metasurface:
\begin{equation}
\displaystyle
\eta_0\aeeoh=\frac{1}{\eta_0} \ammoh = i\frac{S}{\omega}.
\label{eq:q2}
\end{equation}
Thus, the effective electric and magnetic polarizabilities of the unit cell normalized to the free-space impedance must be equal (corresponding to balanced electric and magnetic properties of the array) and purely imaginary (corresponding to a resonance of the inclusions in the array). This is a very important fact meaning that in order to totally absorb incident electromagnetic radiation, the structure must possess equally significant electric and magnetic properties, and that \emph{both} electric and magnetic dipolar responses of the array must resonate at the same frequency.

From Eqs.~(\ref{eq:q1}) the condition of perfect matching of the absorber (no reflections) reads
\begin{equation}
\displaystyle
\eta_0\aeeoh=\frac{1}{\eta_0} \ammoh .
\label{matching}
\end{equation}
Here we will show that it is possible to realize a lossy and resonant array of inclusions so that the condition of full absorption (\ref{eq:q2}) is satisfied at one frequency, but the condition of zero reflection (\ref{matching}) holds in an ultra-wide frequency range, although both polarizabilities are frequency dispersive.

For the design of unit cell topologies it is more convenient to work with the individual polarizabilities of the inclusions in free space $\aee$ and $\amm$, which are related to the effective ones as \cite{Teemu}
\begin{equation}
\displaystyle
\frac{1}{\eta_0\aee}=\frac{1}{\eta_0\aeeoh}+\frac{\be}{\eta_0}, \quad \frac{1}{\amm / \eta_0}=\frac{1}{\ammoh/\eta_0}+\frac{\be}{\eta_0},
\label{eq:q3}
\end{equation}
where $\be$ is the interaction constant of the infinite periodic array of electric dipoles. Using the known expression for the interaction constant  \citep{tretyakov}, we can find the required individual polarizabilities of the inclusions of a symmetric absorber:
\begin{equation}
\displaystyle
\frac{1}{\eta_0\aee}=\frac{1}{\amm / \eta_0}  = \Re\left(\frac{\be}{\eta_0}\right)-i \frac{\omega k^2}{6\pi}-i \frac{\omega}{2S},
\label{eq:q4}
\end{equation}
where $k$ is the free-space wave number. As one can see from (\ref{eq:q4}), the required normalized individual electric and magnetic polarizabilities of each unit cell must be equal. Moreover, the complex quantity on the right-hand side of (\ref{eq:q4}) reveals that the frequency at which the inclusions resonate in the absorbing array differs from the resonance frequency of one single inclusion in free space.

\begin{figure}[h]
\centering
 \subfigure[]{
   \epsfig{file=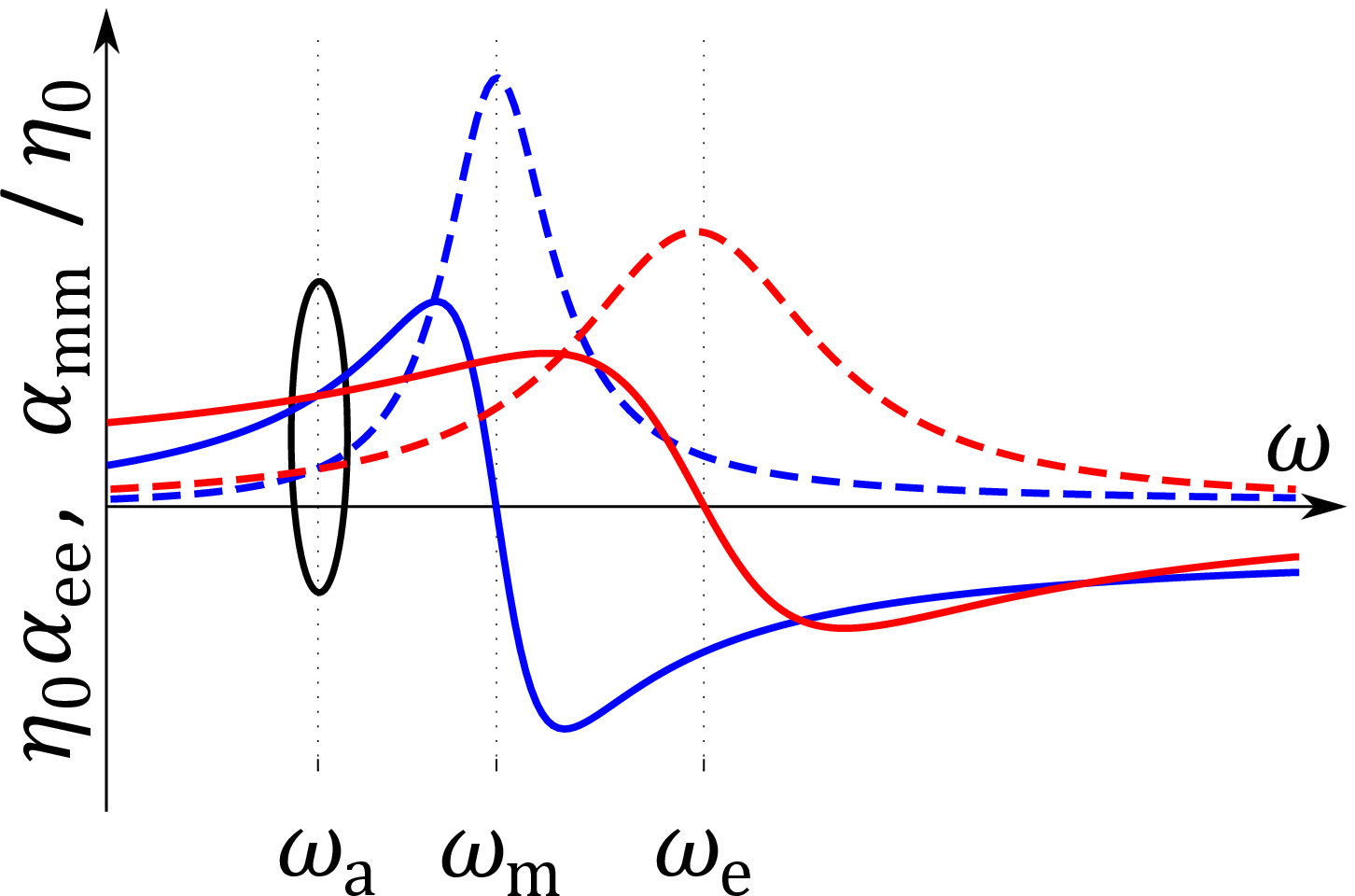, width=0.45\columnwidth}
   \label{ris:fig2a} }
  \subfigure[]{
   \epsfig{file=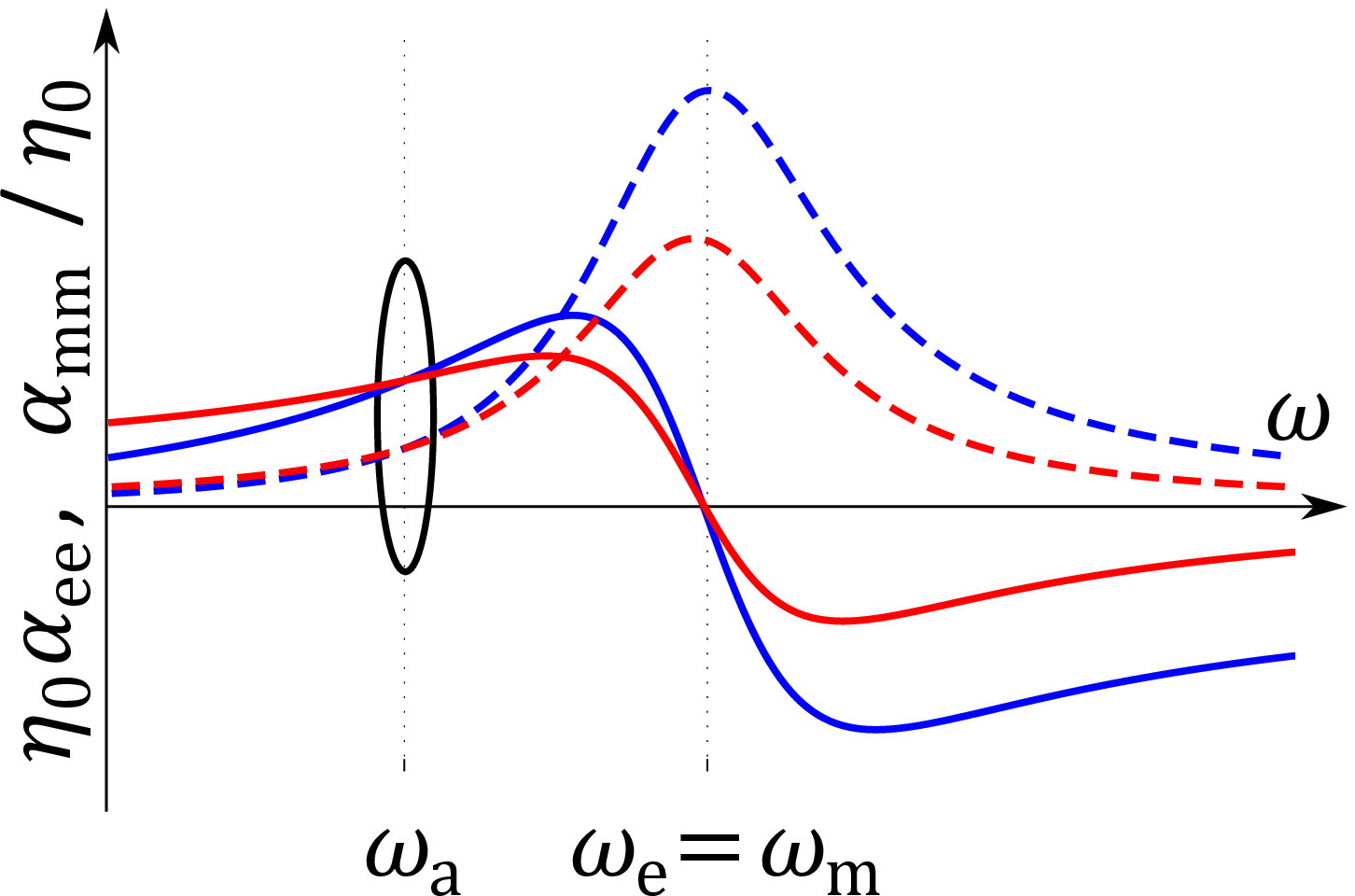, width=0.45\columnwidth}
   \label{ris:fig2b} }
\caption{(color online) Illustration of two absorption regimes in a metasurface with  unit cells containing electrically and magnetically polarizable inclusions resonating (a) at different frequencies and (b) at the same frequency. Red and blue lines depict, respectively, normalized electric and magnetic polarizabilities. Solid and dashed lines show the real and imaginary parts of the polarizabilities, respectively. Calculation parameters: (a) $\omega_{\rm m}=0.86\omega_{\rm e}$, $A=1.28\omega_{\rm e}^2$, $B=1$, $\gamma_{\rm e}=0.2\omega_{\rm e}$, $\gamma_{\rm m}=0.1\omega_{\rm m}$, $\omega_{\rm a}=0.74\omega_{\rm e}$; (b) $\omega_{\rm e}=\omega_{\rm m}$, $A=\omega_{\rm e}^2$, $B=1.56$, $\gamma_{\rm e}=0.2\omega_{\rm e}$, $\gamma_{\rm m}=0.2\omega_{\rm m}$, $\omega_{\rm a}=0.8\omega_{\rm e}$.}
\label{ris:fig2}
\end{figure}

To the best of our knowledge, all known thin absorbers can be divided into two kinds on the basis of the method of their matching. Absorbers of the first kind consist of unit cells, each of which supports electric and magnetic dipole modes resonating at {\it different} frequencies $\omega_{\rm e}$ and $\omega_{\rm m}$, respectively. The unit cell may consist of one single or two separated inclusions (for example, an electrically polarizable straight piece of metal wire and a magnetically polarizable split-ring resonator).
According to Eqs.~(\ref{eq:q4}), at the perfect-absorption frequency $\omega_{\rm a}$, the inclusions should have equal complex polarizabilities. Fig.~\ref{ris:fig2a} corresponds to the situation when the electric and magnetic dipole modes of the inclusions resonate at different frequencies but are tuned so that at the design frequency $\omega_{\rm a}$ the polarizabilities (both real and imaginary parts) are equal. 
Here we model the polarizabilities using the conventional Lorentz 
dispersion model which near the resonance adequately describes electric and magnetic dipolar response of small inclusions of arbitrary nature:
\begin{equation}
\alpha_{\rm ee}={A\over \omega_{\rm e}^2-\omega^2-i\omega \gamma_{\rm e}},
\label{eq:q6}
\end{equation}
\begin{equation}
\alpha_{\rm mm}={B\omega^2 \over \omega_{\rm m}^2-\omega^2-i\omega \gamma_{\rm m}}.
\label{eq:q7}
\end{equation}
Obviously, the electric and magnetic responses of the unit cell tuned for one specific frequency become unbalanced at other frequencies. This inevitably yields non-zero reflectivity of the absorber at frequencies outside of the operational band, since condition (\ref{matching}) is satisfied only at one frequency. Typical absorbers possessing such properties have been reported in Refs.~[\onlinecite{ref1,ref2,ref4}].

Absorbers of the second kind have a unit cell comprising one single or two separate inclusions which support electric and magnetic dipole modes resonating at the {\it same} frequency $\omega_{\rm e}=\omega_{\rm m}$. 
One can show based on dispersion rules (\ref{eq:q6}) and (\ref{eq:q7}) that it is possible to make the two polarizabilities equal only at one single frequency, while at other frequencies the different polarizabilities result in undesired reflection from the absorbing structure [see Fig.~\ref{ris:fig2b}].
The physical reason for this is that distributions of conduction and/or polarization currents in the inclusion volume are different for different modes. Moreover, the two loss factors inevitably have different dispersions because the radiation damping factor for a small electric dipole behaves as $\omega^2$ while for a magnetic dipole the dependence is $\omega^4$. Therefore, the loss factors and their frequency dependence are inevitably different. Thus, the dispersion curves of two different modes in principal cannot be identical.
We can conclude that also in thin absorbers of this kind at least moderate reflection inevitably appears outside of the absorption band, see examples in \cite{7padilla,bilotti}.
Note that the problem of the unbalanced dispersion curves of the electric and magnetic modes is inherent in all known resonators used for wave manipulations, for example, core-shell nanoparticles \cite{core-shell1,core-shell2} and dielectric cubes \cite{arxiv}.

To overcome the described drawbacks, we propose to use inclusions designed in such a manner that \emph{both} electric and magnetic responses are created by excitation of \emph{the same resonant mode}. The unique feature of the proposed new kind of metasheets is that they possesses zero reflectivity over an ultra-wide frequency range and are totally transparent (without polarization conversion) outside of the resonant band. The only known alternative approach to  realization of all-frequency matched absorbers is a theoretical idea of using a slab made of a lossy material with equal values of the relative permittivity and permeability in a wide frequency range \cite{Padilla}. Besides the fact that such media do not exist, even from the conceptual point of view this would require a slab of infinite thickness in order to ensure zero transmission at the frequency of absorption.

\section{Total absorption in an array of helices}

Based on the above considerations, we propose to use a resonant mode of a single inclusion which is coupled to both electric and magnetic fields. The induced current distribution of this resonant mode should be such that both electric and magnetic moments are excited and can be tuned to the desired balance. These properties are typical for bianisotropic elements. However, as it was discussed, to behave as a symmetric single-layer perfect absorber, the structure must be \emph{not}  bianisotropic.  
In order to surmount this obstacle, we propose an unprecedented route: using bianisotropic inclusions on the level of the unit cell but arranging the inclusions in the array so that the bianisotropy is compensated on the level of the entire array. We achieve this by alternating bianisotropic inclusions of two sorts in the array. These two sorts differ only by the sign of the electromagnetic coupling  parameter, therefore, combination of them yields bianisotropy compensation. Thus, keeping zero electromagnetic coupling within the array, at the same time we create coupled electric and magnetic dipole moments in each unit cell resonating exactly at the same frequency and having the same loss factors. Since the inclusions of the two sorts possess identical dipole moments, no splitting of the resonance frequency of the structure occurs.

A possibility to engineer helices so that both amplitude and loss factors in the dispersion rules (\ref{eq:q6}) and (\ref{eq:q7}) are equal can be seen from the antenna model of canonical wire helices \cite{antenna_model}. Equal amplitudes can be ensured simply by proper choosing the helix dimensions. The loss factors $\gamma_{\rm e}$ and $\gamma_{\rm m}$ of the helix are identical because both electric and magnetic polarizabilities depend on the \emph{sum} of the radiation resistances of a small electric dipole and a small magnetic dipole excited in the helix \cite{antenna_model}. 
Thus, the electric and magnetic polarizabilities of the inclusions have nearly identical dispersions and the proposed array of the inclusions acts as a Huygens' surface in a very wide frequency range. Deviations occur only far from the resonance, where the array is anyway very weakly excited and reflections are negligible. 

Although in this paper we use the idea of using bianisotropic elements for wide-band matching absorbers, it can be utilized also for various other devices, for example, transmit-arrays \cite{Huygens1,Huygens2,Huygens3}.
In this paper we apply and experimentally validate our concept of all-frequency-matched Huygens' metasurfaces to absorbers in the microwave frequency range. However, the concept is generic and can be applied over the entire electromagnetic spectrum. 

\begin{figure}[h]
\centering
 \subfigure[]{
   \epsfig{file=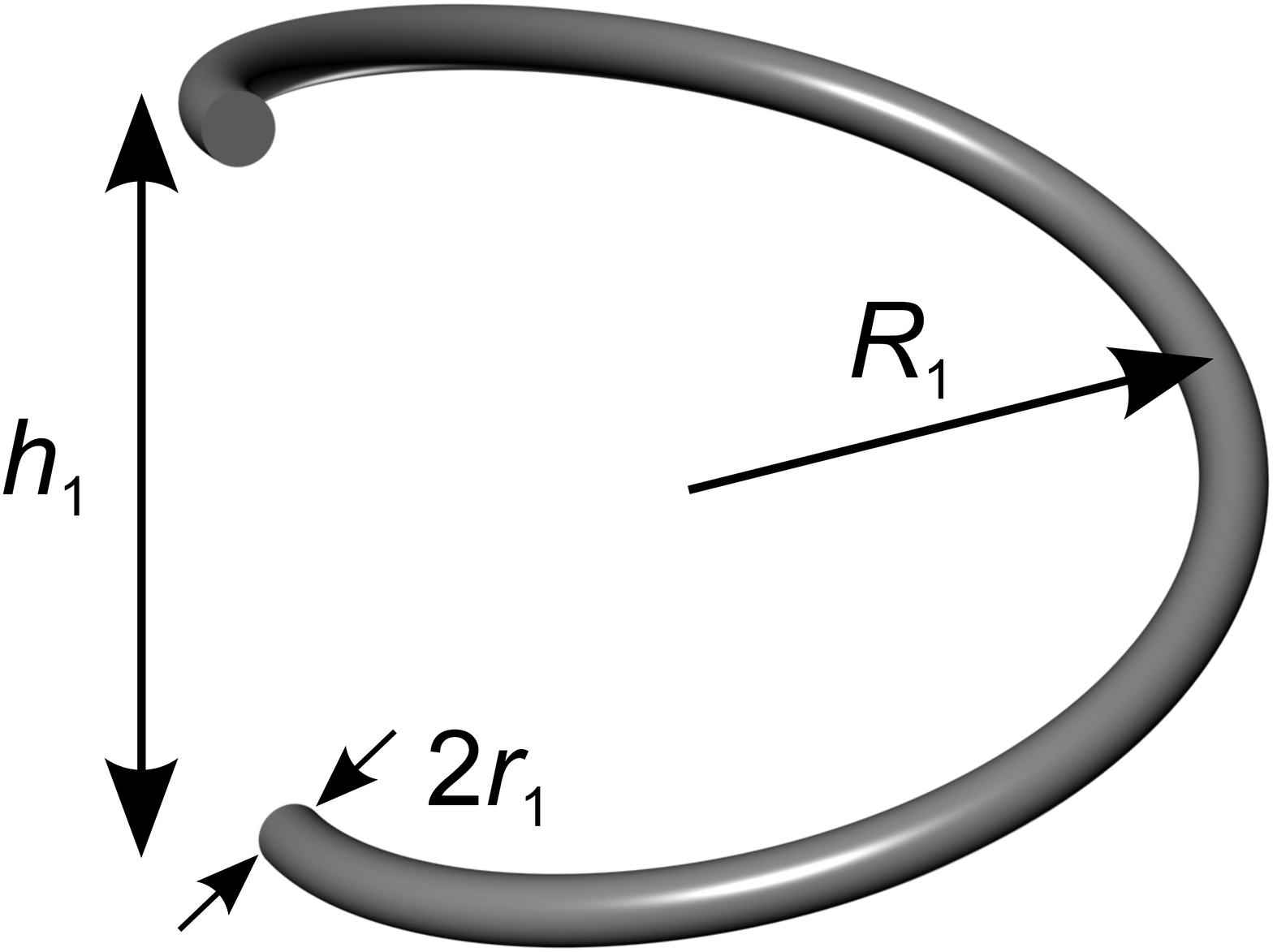, width=0.45\columnwidth}
   \label{ris:fig3a} }
  \subfigure[]{
   \epsfig{file=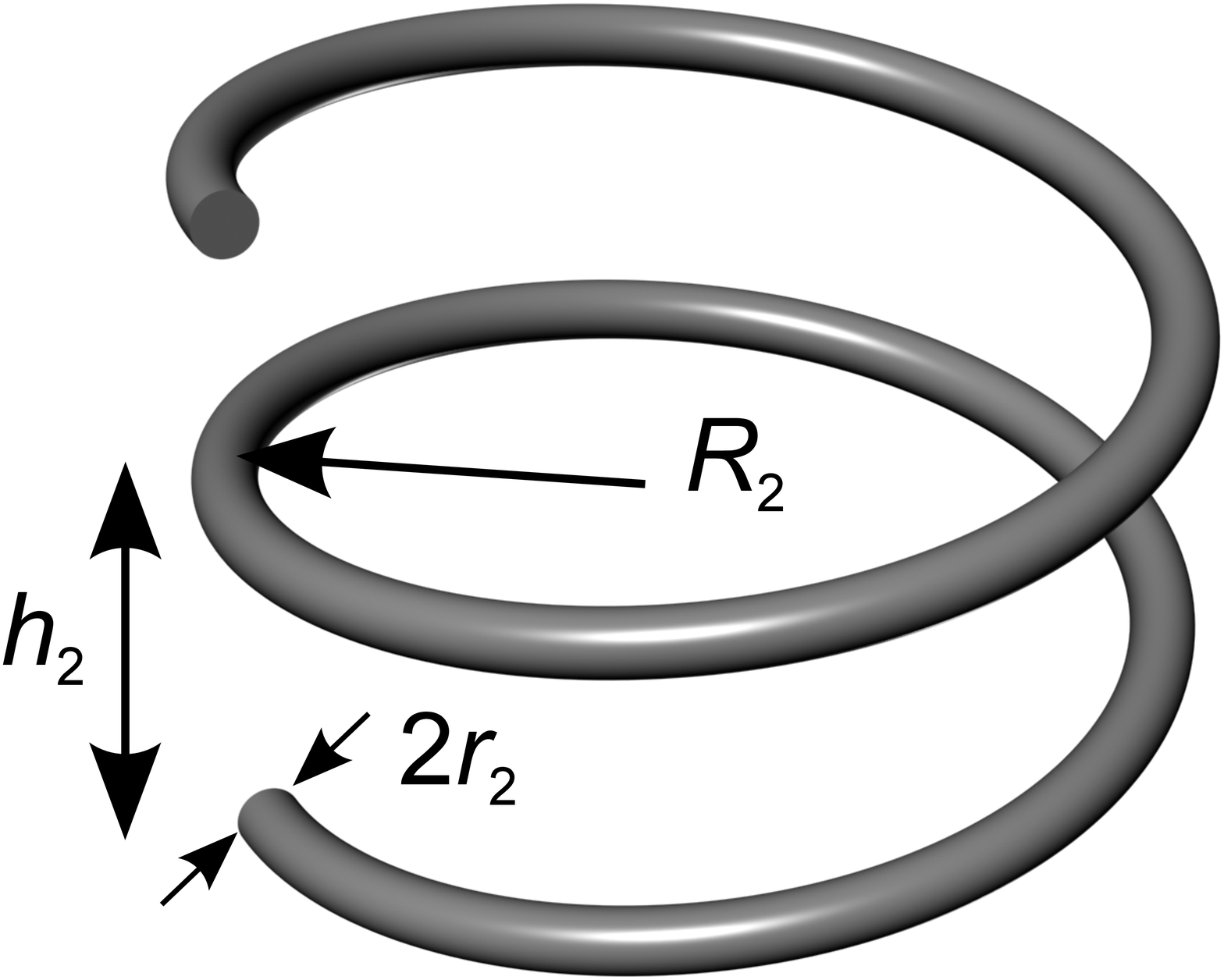, width=0.45\columnwidth}
   \label{ris:fig3b} }\\
     \subfigure[]{
   \epsfig{file=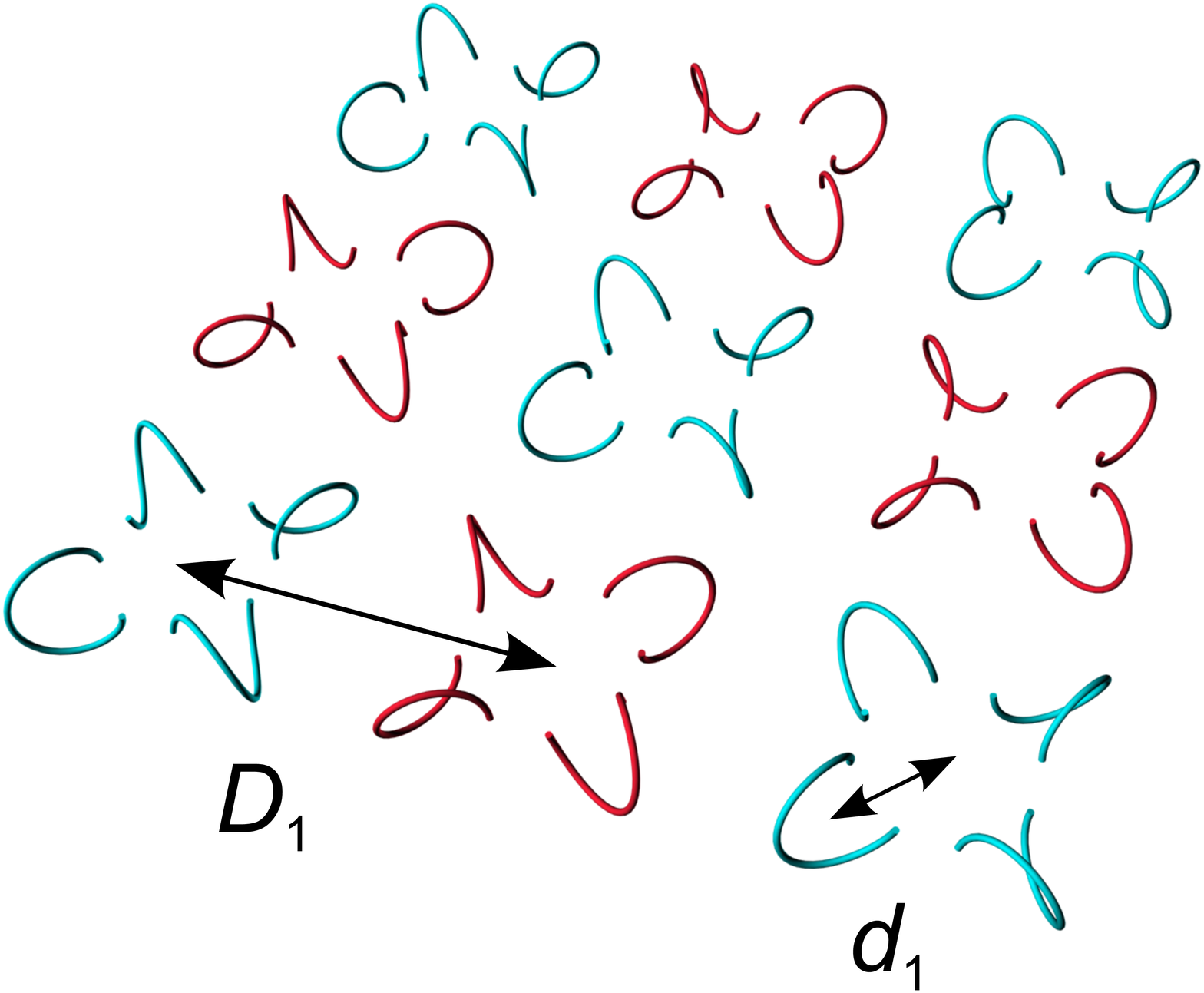, width=0.45\columnwidth}
   \label{ris:fig3c} }
     \subfigure[]{
   \epsfig{file=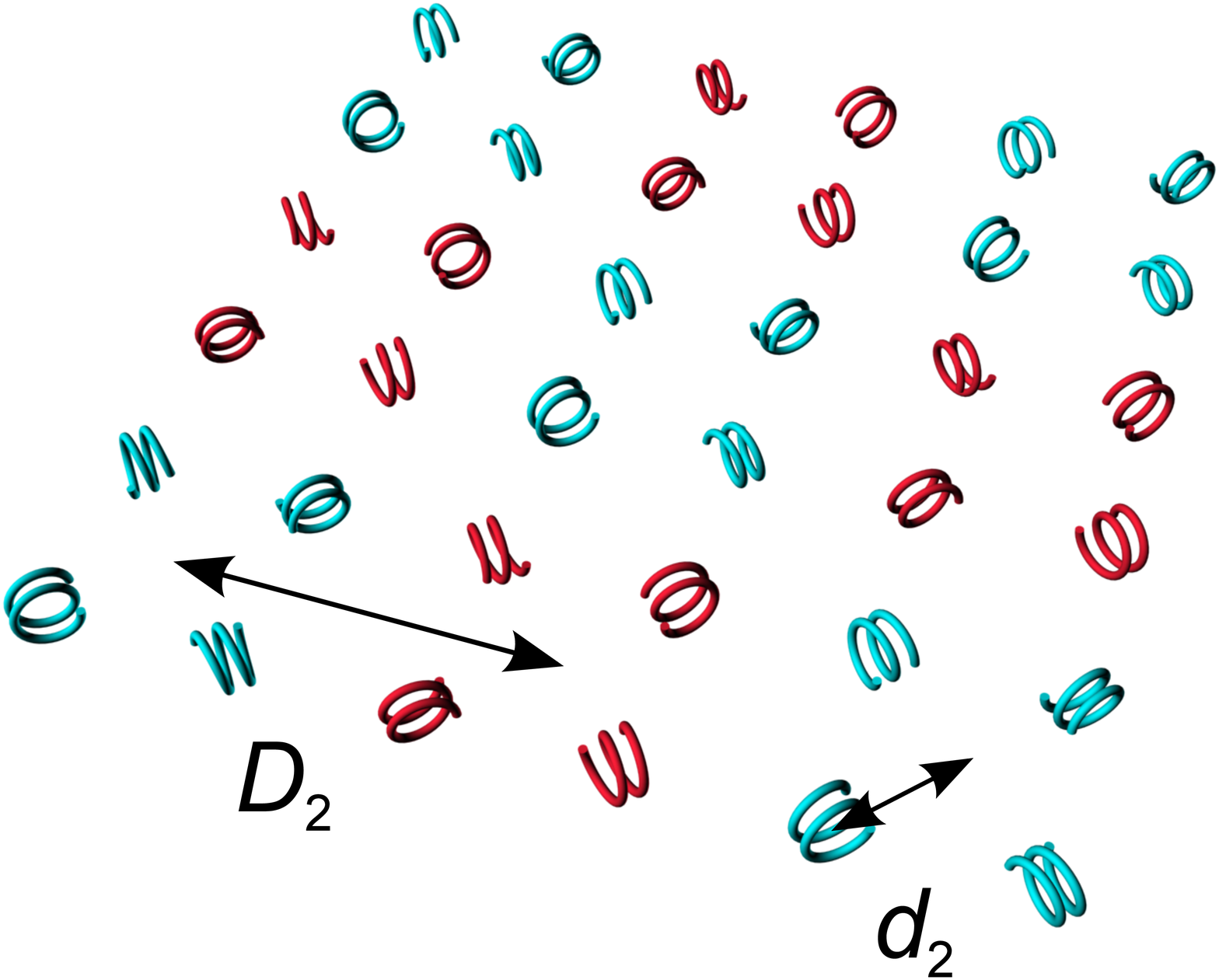, width=0.45\columnwidth}
   \label{ris:fig3d} }
\caption{(color online) (a) Single-turn and (b) double-turn helical inclusions. (c) Arrangement of the single-turn and (d) double-turn helical inclusions in the arrays. Blue and red colors denote right- and left-handed helices, respectively.}
\label{ris:fig3}
\end{figure}

Here we propose to use chiral \cite{serdukov} bianisotropic elements as inclusions of an absorber, although one can realize similar scenarios based on other types of bianisotropic elements. There are many different topologies of chiral elements: for example, helices \cite{ozbay}, chiral split ring resonators \cite{chiral}, cut-wire stacks \cite{alu}. In view of simplicity of the design and realization we utilize smooth helical inclusions \cite{conference2} (see Figs.~\ref{ris:fig3a} and \ref{ris:fig3b}). Arrays of such inclusions operating at infrared frequencies can be manufactured based on fabrication technologies reported in Ref.~[\onlinecite{fabrication1,fabrication2}].
We consider single-turn and double-turn wire helices as two fundamental classes of helical elements. Other helices with a higher odd (even) number of loops have properties similar to those of a single-turn (double-turn) helix \cite{samofalov2}.
Due to coupled electric and magnetic moments, single-turn and double-turn helices have sub-wavelength dimensions: $\lambda/7$ and $\lambda/15$ at the operating frequency, respectively. This ensures homogenization of the layer response and decreases its thickness. Compensation of chiral bianisotropy in an absorber can be achieved by alternating left- and right-handed helices.

We position the helices in the $xy$-plane in such a manner that the array has the fourfold rotational symmetry about the propagation $z$-axis and, therefore, is polarization insensitive. 
An optimal arrangement of helices in the array was considered in Ref.~[\onlinecite{conference1}] and depicted in Figs.~\ref{ris:fig3c} and \ref{ris:fig3d}. The super-unit-cell of the array consists of four blocks of helices.
The blocks comprising four helices of specific handedness are staggered in the array. The size of such a structural block in the case of single-turn helices is $D_1=0.6\lambda=56$~mm and in the case of double-turn helices is $D_2=0.4\lambda=43$~mm. The single- and double-turn helices are located, respectively, at distances $d_1=0.2\lambda=19.8$~mm and $d_2=0.2\lambda=15.1$~mm from the centres of the corresponding blocks.

Individual electric $\aee$ and magnetic $\amm$ polarizabilities of the helices can be made equal according to requirements (\ref{eq:q4}) by varying the helices geometry. Using analytical expressions \cite{antenna_model,samofalov2,sks} and the method for extracting polarizabilities introduced in Ref.~[\onlinecite{polarizability}], we have optimized the dimensions of the inclusions. For single-turn helices, the loop radius is $R_1=7.2$~mm, the height of the loop is $h_1=11.3$~mm, and the wire radius $r_1=0.1$~mm. The corresponding parameters of the double-turn helices are as follows: $R_2=3.3$~mm, $h_2=2.3$~mm, and $r_2=0.25$~mm. 
The design frequency 3~GHz corresponds to the half-wavelength resonance frequency of the helices. 
The helical inclusions are embedded in a plastic foam substrate (for mechanical support) with $\epsilon=1.03$ and the thickness 14.4~mm. 

\begin{figure}[h]
\centering
   \epsfig{file=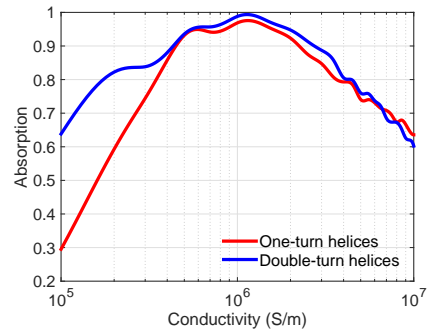, width=0.7\columnwidth}
\caption{(color online) Absorption coefficient versus conductivity of the inclusions material at the operating frequency of 3~GHz.}
\label{ris:fig4}
\end{figure}

Full-absorption regime is accomplished in the array with a proper level of dissipative loss in the helical inclusions. The absorption level versus conductivity of the material of the inclusions obtained with a commercial electromagnetic software \cite{ANSYS} is plotted in Fig.~\ref{ris:fig4}.
It is seen that larger or smaller than the optimum levels of dissipative loss in the inclusions degrade the performance of the absorbers. In our design we utilize helices made of nichrome NiCr60/15 with the conductivity $10^6$~S/m, which approximately ensures the required level of dissipation loss (see Fig.~\ref{ris:fig4}). 

\begin{figure}[h]
\centering
 \subfigure[]{
  \epsfig{file=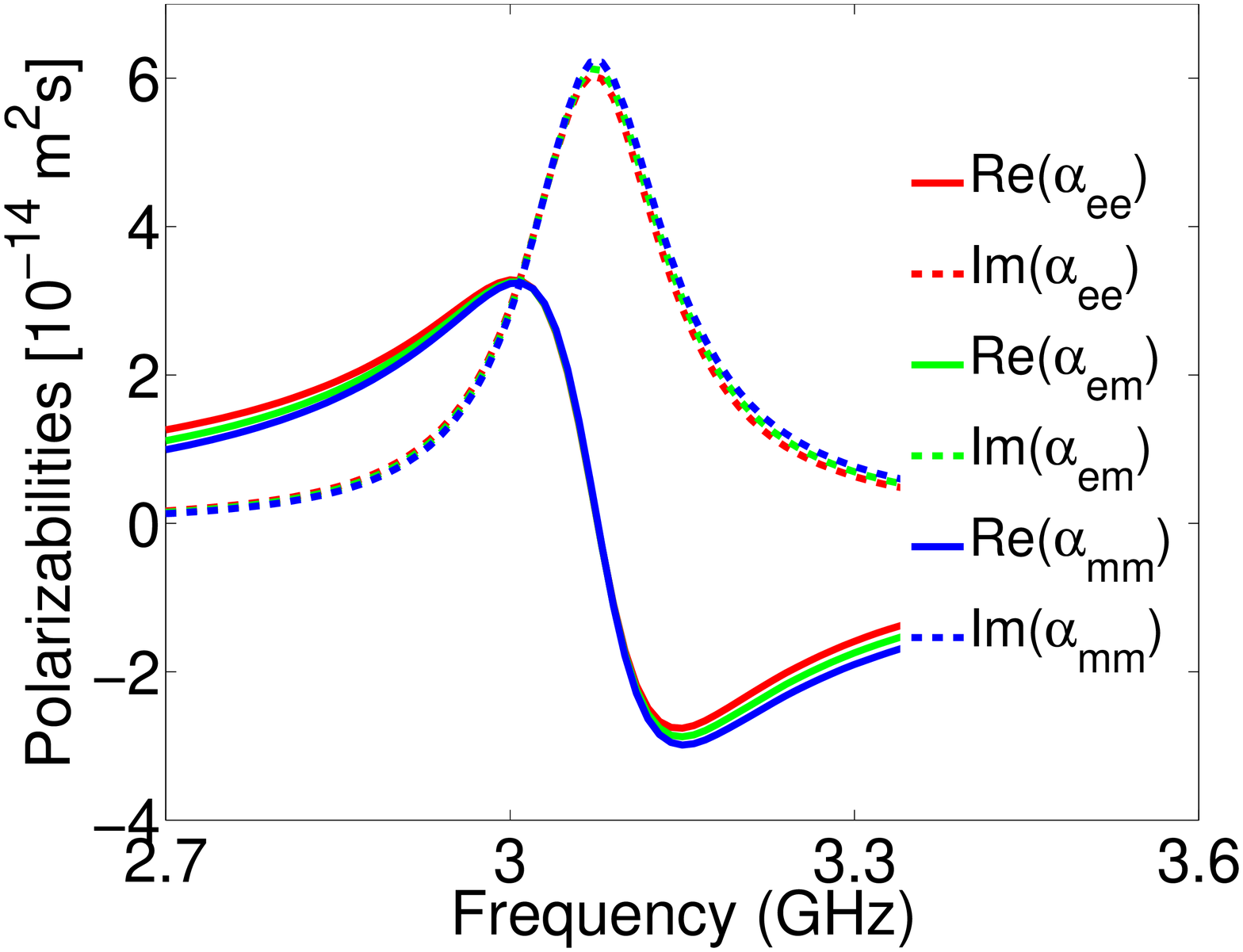, width=0.45\columnwidth}
   \label{ris:fig5a} }
  \subfigure[]{
   \epsfig{file=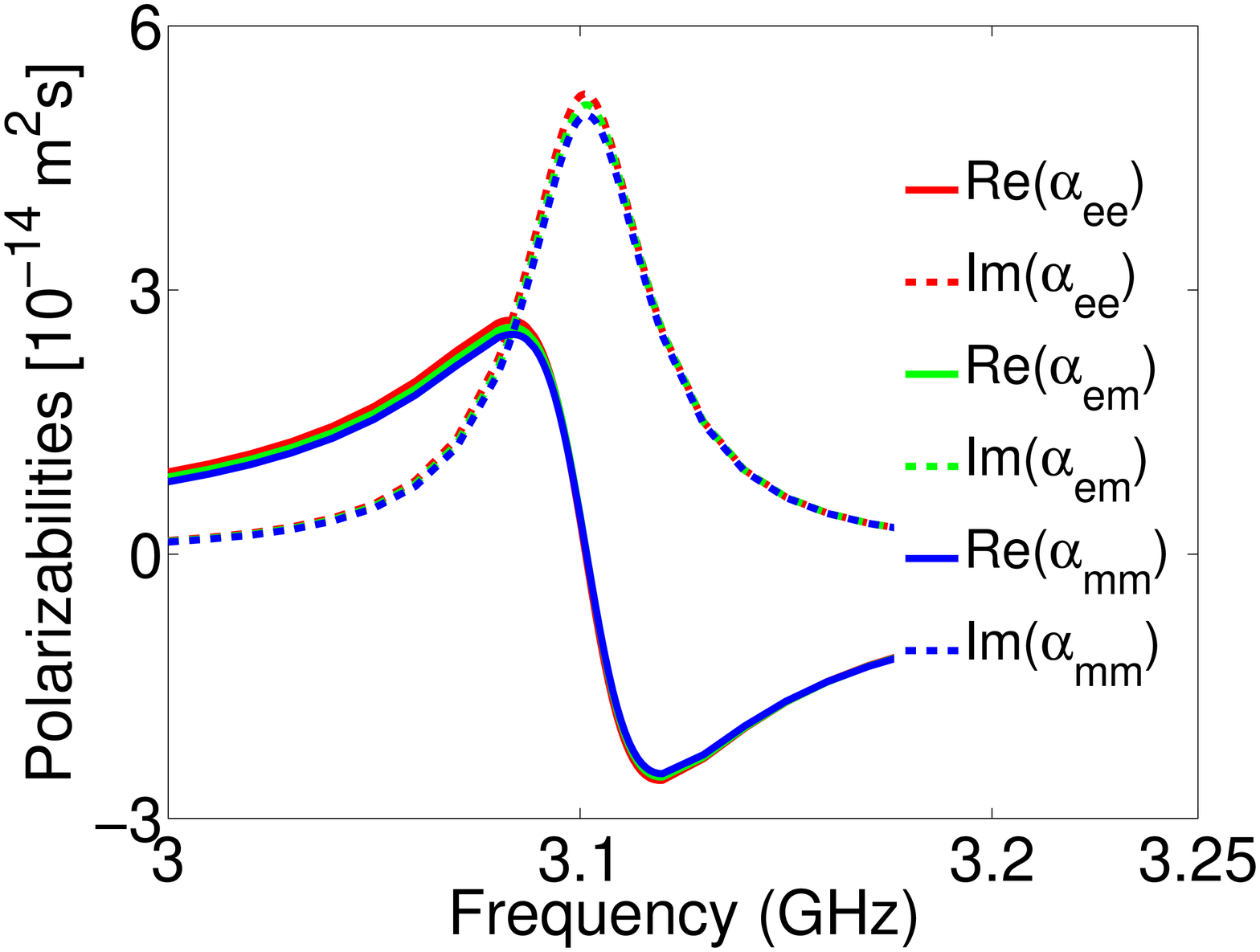, width=0.47\columnwidth}
   \label{ris:fig5b} }
\caption{(color online) Individual normalized polarizabilities of the designed (a) single- and (b) double-turn helical inclusions.}
\label{ris:fig5}
\end{figure}

The individual polarizabilities of the single-turn and double-turn helices satisfying requirements of total absorption~(\ref{eq:q4}) are shown in Fig.~\ref{ris:fig5}. 
The non-zero electromagnetic polarizability $\alpha_{\rm em}$ of a single inclusion is compensated in the array of the left- and right-handed helices. 
It is seen that the electric and magnetic polarizabilities  have nearly identical frequency dispersions, meaning that they possess desired electromagnetic properties of wide-band Huygens' sources. This unique feature distinguishes our absorber from known designs whose structural elements possess spectrally different dispersions of the electric and magnetic modes [see Figs.~\ref{ris:fig2a} and \ref{ris:fig2b}].

\begin{figure}[h]
\centering
 \subfigure[]{
   \epsfig{file=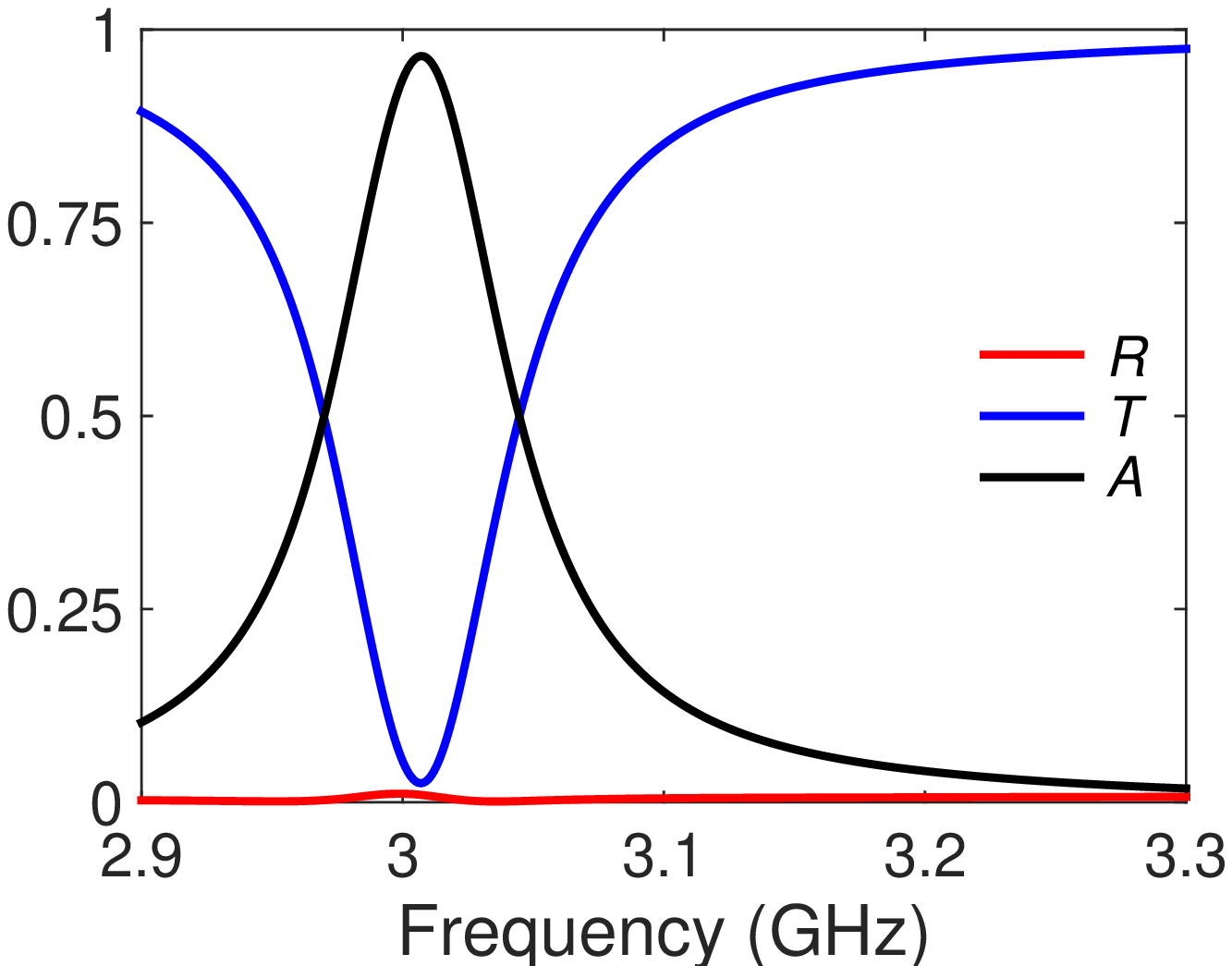, width=0.45\columnwidth}
   \label{ris:fig6a} }
  \subfigure[]{
  \epsfig{file=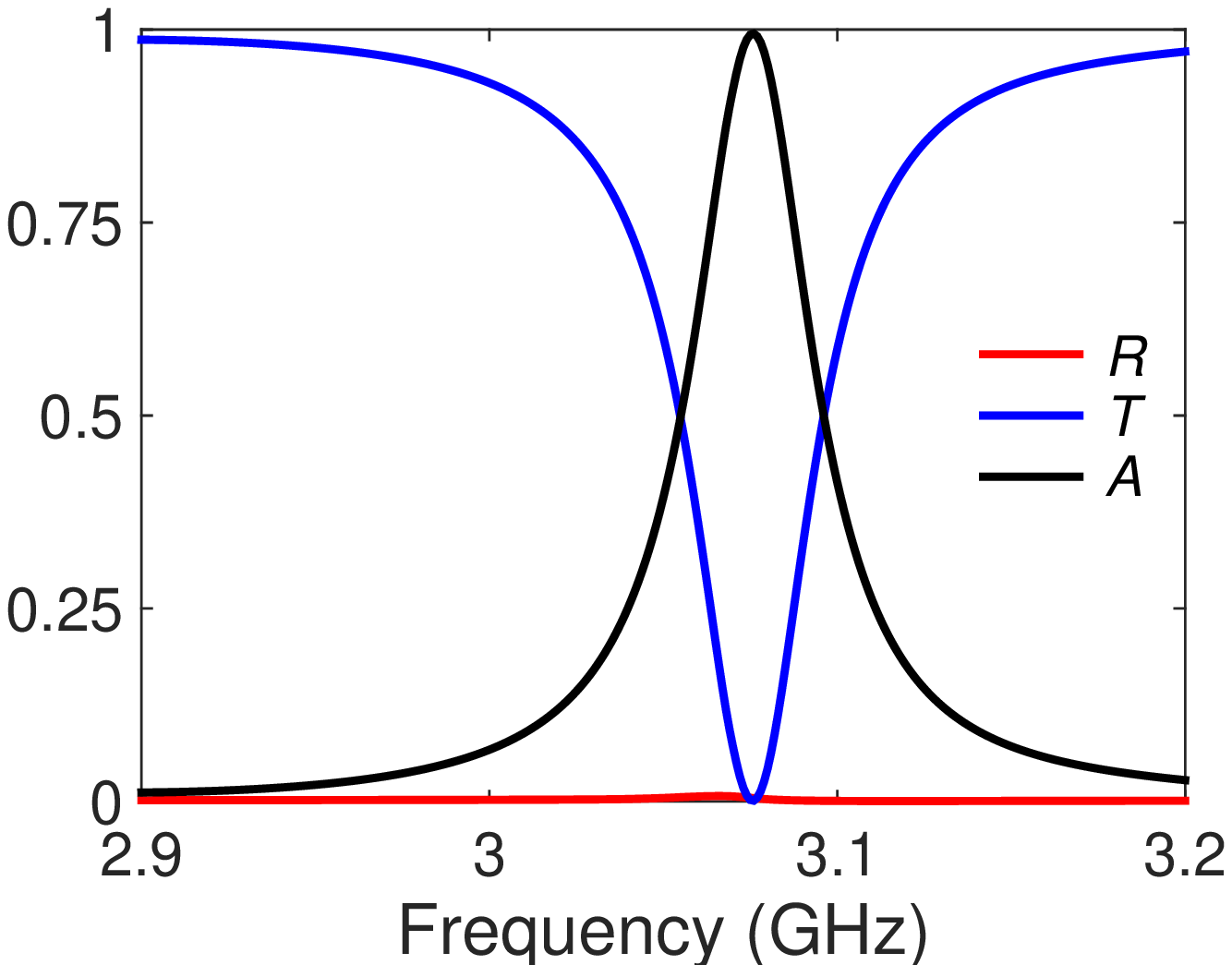, width=0.45\columnwidth}
   \label{ris:fig6b} }\\
    \subfigure[]{
   \epsfig{file=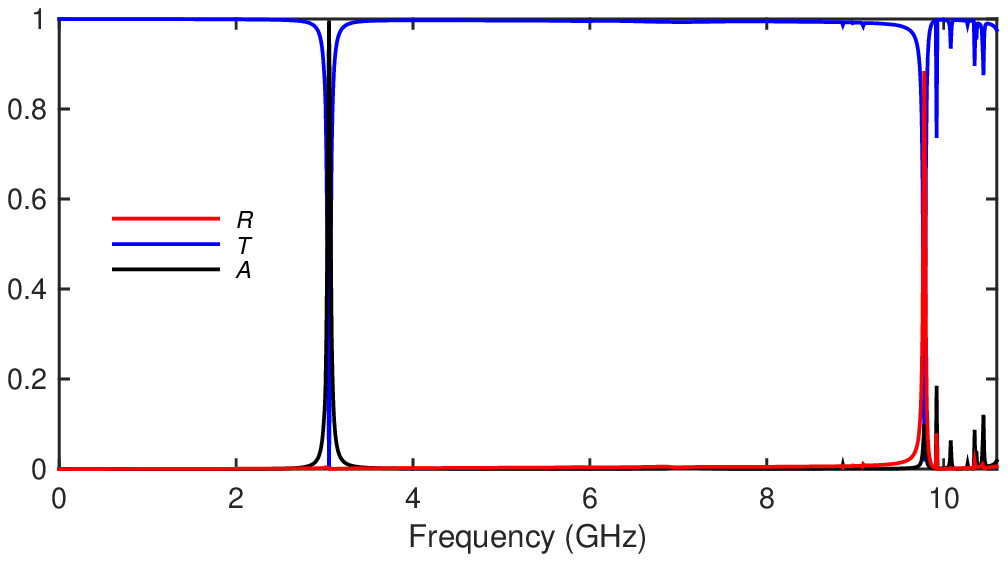, width=0.95\columnwidth}
   \label{ris:fig6c} }
\caption{(color online) Simulated power reflection $R$, transmission $T$, and absorption $A$ coefficients for infinite arrays of (a) single- and (b) double-turn helical inclusions. (c) Response of the double-turn helix array in a wide frequency range.}
\label{ris:fig6}
\end{figure}

The reflection, transmission and absorption coefficients for infinite arrays of the described single- and double-turn helices are plotted in Figs.~\ref{ris:fig6a} and \ref{ris:fig6b}.
One can see that at the resonance frequency the designed metasurfaces based on single- and double-turn helices absorb 96.5\% and 99.9\% of the incident power, respectively. 
The absorbers, due to their symmetrical geometry, identically operate with waves impinging on either of their side. 
It can be seen that both inside and outside of the absorption band the reflection coefficient is practically zero for both metasurfaces. 
In order to highlight the unique functionality of our absorbers, we illustrate in Fig.~\ref{ris:fig6c} the broadband response of the metasurface based on double-turn helices. While the transmission coefficient has a resonance of full absorption at about 3 GHz, the reflection coefficient is  not distinguishable from zero in an ultra-wide range from MHz frequencies to about 10 GHz, where the circumference of one turn of the helix becomes comparable with the wavelength and the first higher-order resonance appears.

\begin{figure}[h]
\centering
   \epsfig{file=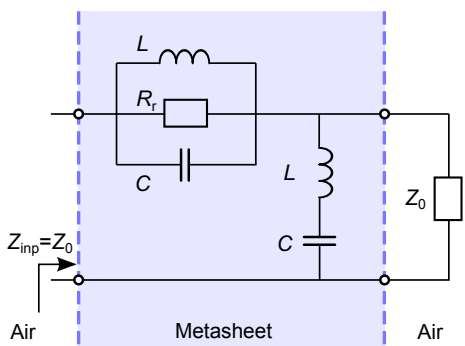, width=0.7\columnwidth}
\caption{An equivalent circuit of the proposed absorbers matched at all frequencies. 
Resistive load $Z_0=377$~$\Omega$ denotes the impedance of free space behind the absorber.}
\label{ris:fig7}
\end{figure}

The remarkable operation of the proposed Huygens' absorbers can be clearly illustrated based on a circuit analogy. The circuit shown in Fig.~\ref{ris:fig7} behaves as the proposed absorber illuminated from one side. 
Indeed, a simple calculation shows that if the inductance $L$, capacitance $C$, and the resistance $R_{\rm r}$ satisfy condition $\displaystyle \sqrt{L/C}=R_{\rm r}=Z_0$, 
the input impedance of the absorber equals to the impedance of free space $Z_0$ \emph{at all frequencies} (which yields zero reflection from the absorber).
In the circuit theory such circuits are known and called constant-resistance networks \cite{zobel,norton}, but it appears that mostly lossless circuits have been studied and used. Although the circuit responds as a resistor, it is dispersive and lossy, and the transmission coefficient  has a resonant behavior.  At the resonance, the parallel branch is short-circuited, and therefore, all the incident power is dissipated in the resistor $R_{\rm r}$ (the metasurface totally absorbs the incident wave). At very low or very high frequencies the series branch is short-circuited, while the parallel branch is an open circuit. Therefore, all the incident power without loss is delivered to the load resistor $Z_0$, which models the free space behind the metasheet (the metasheet becomes invisible for electromagnetic radiation at these frequencies). 

\begin{figure}[h]
\centering
 \subfigure[]{
   \epsfig{file=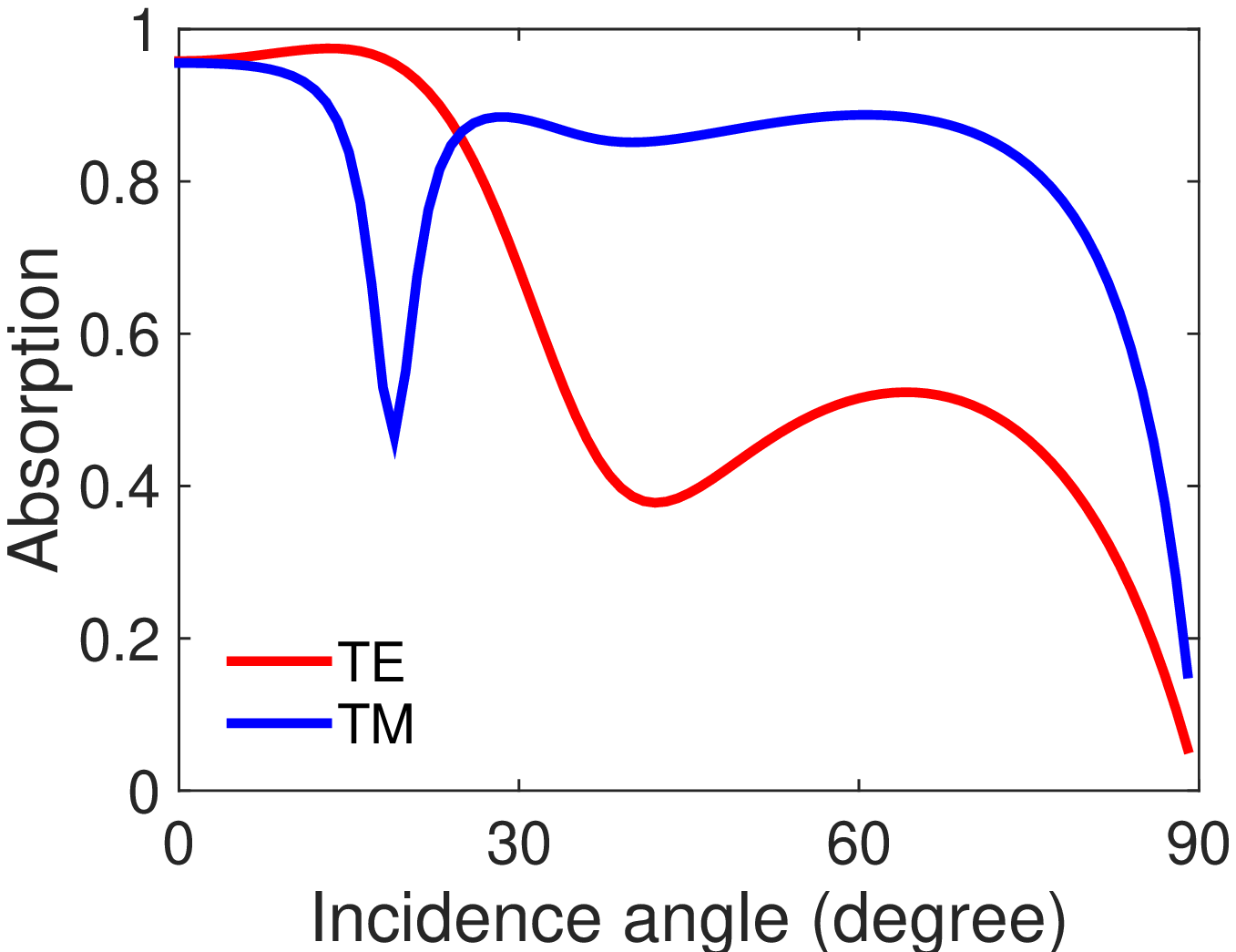, width=0.45\columnwidth}
   \label{ris:fig8a} }
  \subfigure[]{
   \epsfig{file=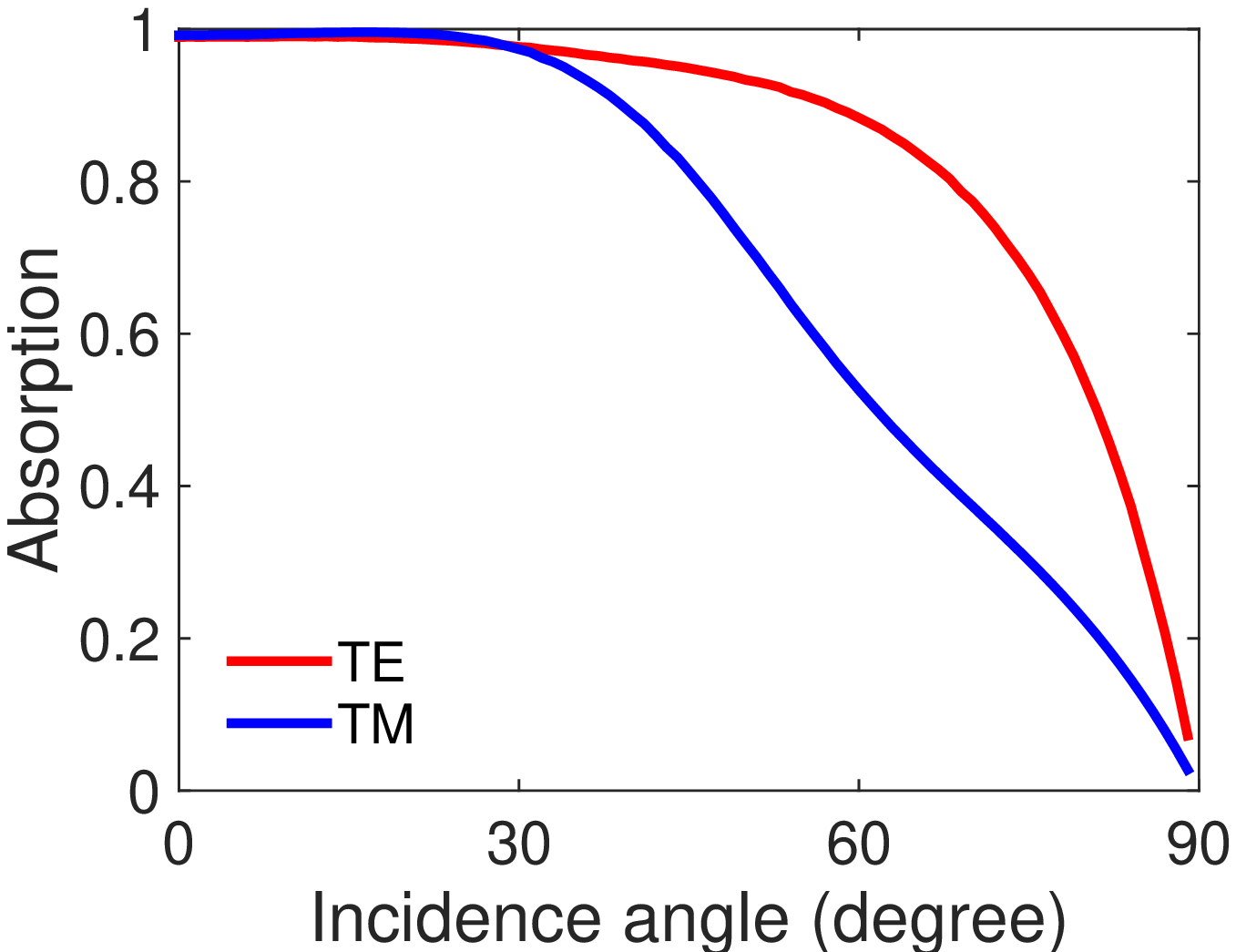, width=0.45\columnwidth}
   \label{ris:fig8b} }
\caption{(color online) The absorption in a metasurface with (a) single-turn and (b) double-turn helical elements versus the incident angle.}
\label{ris:fig8}
\end{figure}

For many applications of absorbers it is of particular importance to absorb normally incident radiation as well as radiation impinging on the structure at oblique angles. 
The angular stability of the proposed absorbers is shown in Fig.~\ref{ris:fig8}. 
It is seen that whereas the metasurface with single-turn helices absorbs nearly perfectly only at the incidence angles close to normal, the metasurface with double-turn helices exhibits impressive angular-stable absorption. This can be explained by the fact that the double-turn helices have more isotropic topologies. For transverse electric (TE) polarized waves illuminating the metasurface with the double-turn helices at angles from 0 to $68^{\circ}$ the absorption level remains above 80\%. Furthermore, for both TE and TM polarizations the absorber operates with efficiency over 95\% at angles from 0 to $35^{\circ}$.  It should be stressed that this high-performance angular stability is achieved in an electrically ultra-thin ($\lambda/15$) structure.

\section{Experimental realization and characterization}
The operation of the two proposed broadband Huygens' metasheet absorbers was verified by conducting measurements in an anechoic chamber based on the free-space method \cite{method}. A microwave signal generator was connected to a transmitting horn antenna, while the signal at the receiving horn antenna was analysed with a microwave receiver. The horn antennas had the apertures of 35~cm~$\times 27$~cm. The samples were located at a distance of 4 meters (about 40$\lambda$) from the transmitting antenna, which approximately secures plane-wave excitation of the samples. The receiving horn antenna was positioned immediately behind the samples.
The fabricated samples were mounted in the hole of a wall made of microwave absorbing material. Transmission measurements were calibrated to the transmission between the horn antennas in the absence of the samples. The reflection measurements were calibrated by using a copper plate with the area equal to that of the samples. In the reflection measurements the transmitting and receiving antennas were each inclined with an angle of about~$3^{\circ}$ with respect to the normal of the samples surface.

\begin{figure}[h]
\centering
 \subfigure[]{
   \epsfig{file=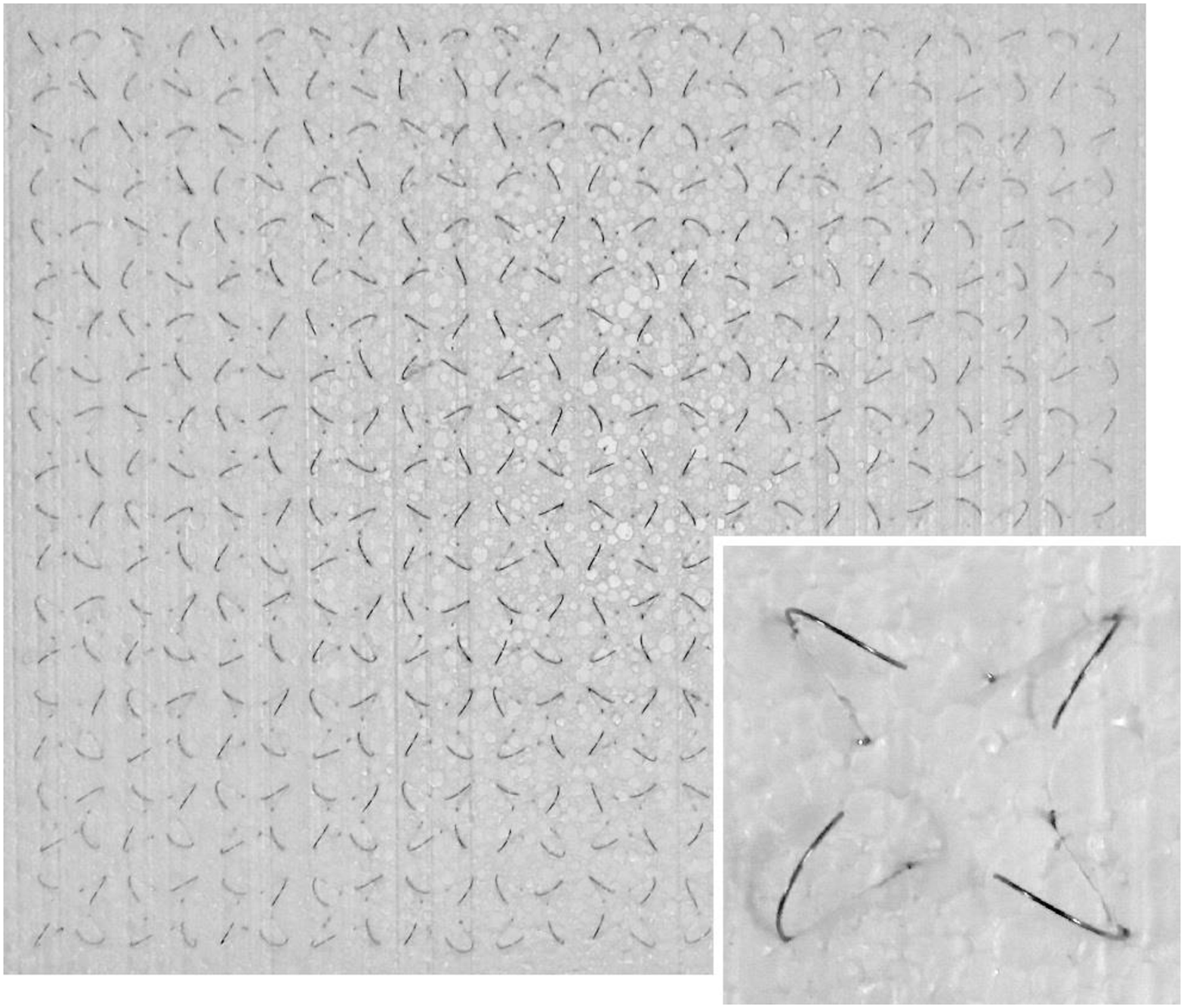, width=0.48\columnwidth}
   \label{ris:fig9a} }
  \subfigure[]{
  \epsfig{file=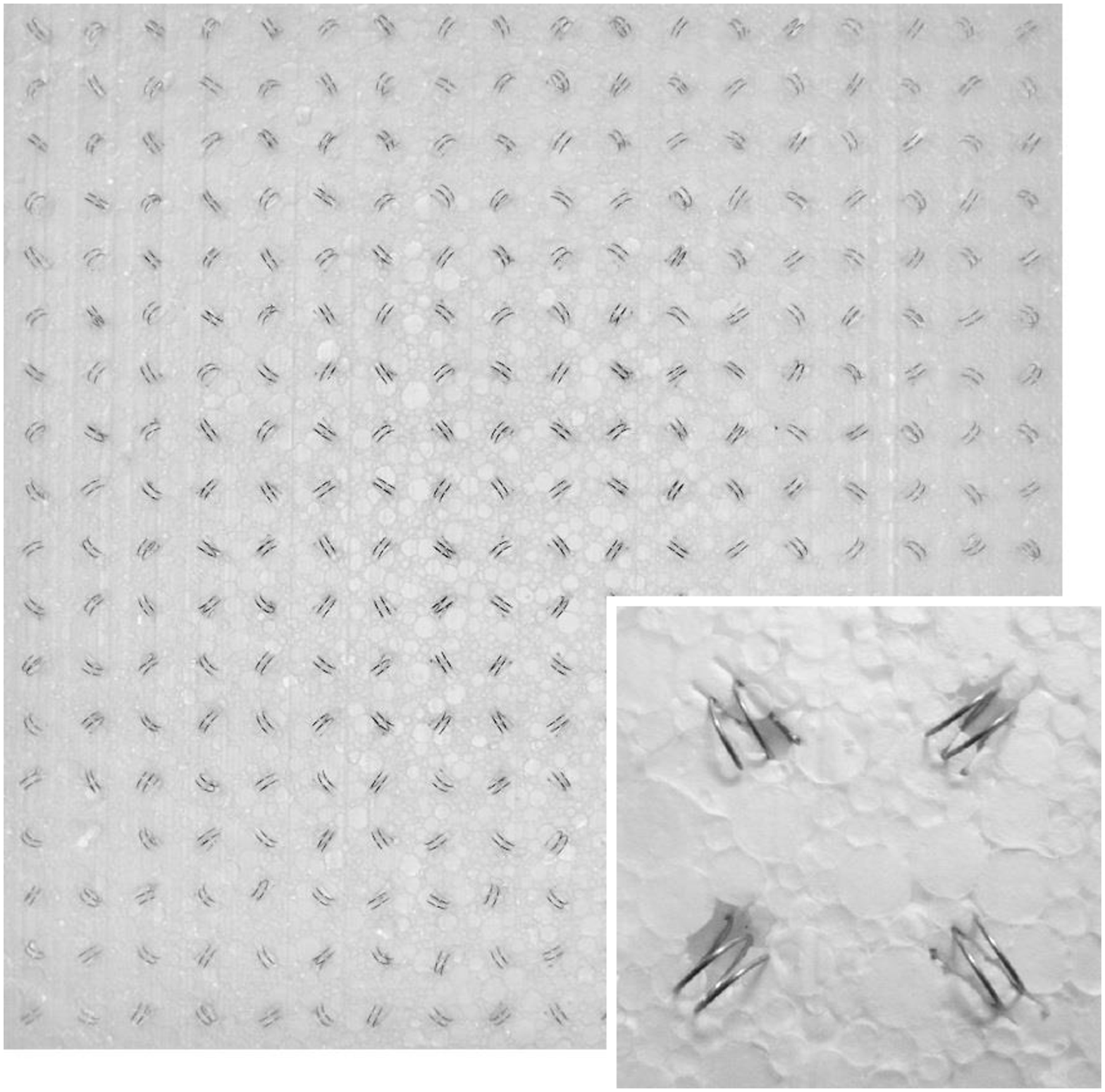, width=0.42\columnwidth}
   \label{ris:fig9b} }
\caption{(color online) Fabricated metasurfaces of (a) single-turn and (b) double-turn helices comprising, respectively, 480 and 324 elements embedded in plastic foam.}
\label{ris:fig9}
\end{figure}

The helical inclusions of the samples were manually manufactured and placed in supporting low-permittivity slab described in the previous section (see Fig.~\ref{ris:fig9}).
Manufacturing single-turn helices with a small wire radius $r_1=0.1$~mm optimized in simulations implies practical difficulties associated with very flexible and unstable wire forming the helix. Therefore, for our measurements we fabricated an array of single-turn helices with the wire radius $r_1=0.25$~mm.

\begin{figure}[h]
\centering
 \subfigure[]{
   \epsfig{file=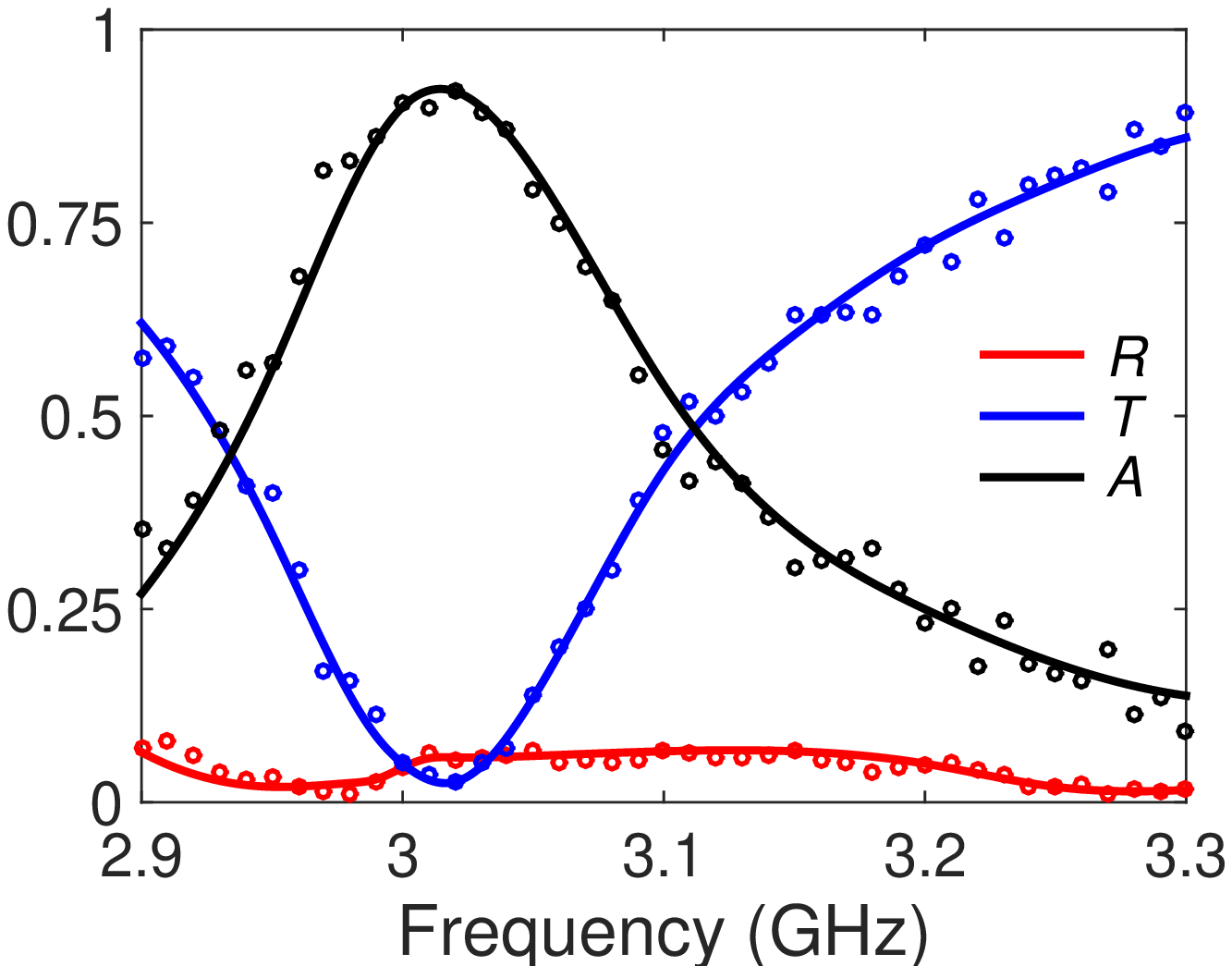, width=0.45\columnwidth}
   \label{ris:fig10a} }
  \subfigure[]{
   \epsfig{file=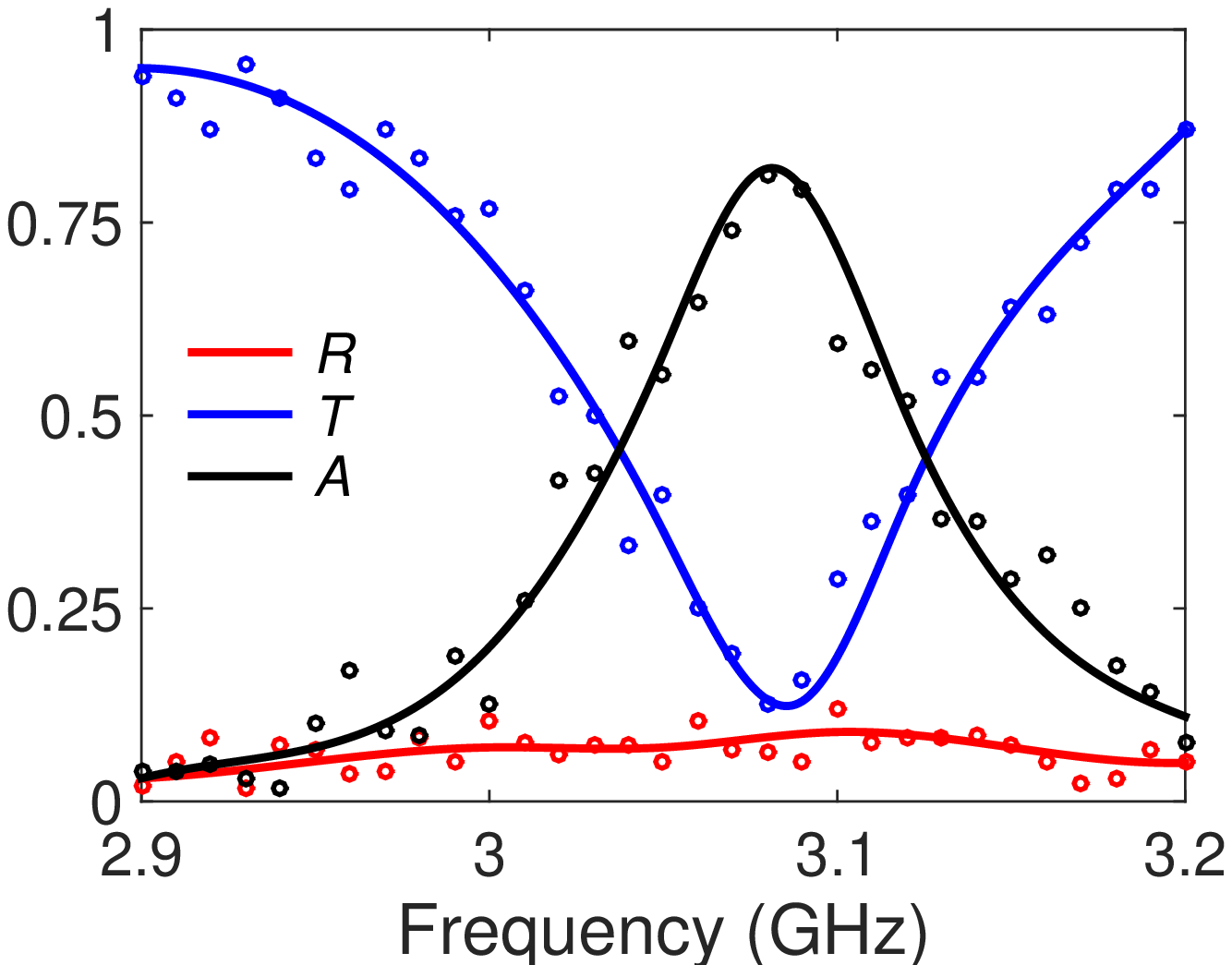, width=0.45\columnwidth}
   \label{ris:fig10b} }
\caption{(color online) Measured reflection, transmission, and absorption coefficients for the fabricated metasurfaces with (a) single- and (b) double-turn helical inclusions. The dots denote measured points. The solid lines denote envelope curves of the measured data.}
\label{ris:fig10}
\end{figure}

The measured power reflection $R$, transmission $T$ and absorption $A$ coefficients of the fabricated samples are depicted in Fig.~\ref{ris:fig10}.
The absorption coefficient was found based on the reflection and transmission data using expression $A=1-R-T$. For an array of single-turn helices the measured peak absorption amounts to 92\%, while for the array of double-turn helices the corresponding quantity achieves 81\%. The experimental curves plotted in Fig.~\ref{ris:fig10} have similar shapes and nearly equal resonance frequencies with the simulated ones (see Fig.~\ref{ris:fig6}). Inaccuracies in manual  fabrication led the manufactured samples to comprise inclusions with slightly varying geometrical dimensions such as the loop radius and the height. As a result, individual inclusions resonate at slightly different frequencies, which inevitably increases the absorption bandwidth of the structures but decreases the absorption peak values. Another reason for the resonance widening is diffuse scattering on array inhomogeneities (slightly varying period, etc.). All these imperfections can be improved with more precise fabrications process of the helical inclusions.
Both the experimental and simulated results manifestly confirm the fact that the proposed absorbers indeed operate as broadband Huygens' metasurfaces, being nearly totally reflection free in and outside of the absorption band.

\section{Discussion and conclusions}

In this paper we have proposed and experimentally tested a novel concept of ultra-thin all-frequencies-matched metasurfaces which produce no reflections in an extremely wide frequency range.
Although here we utilized these metasurfaces for designing absorbers, they can be perfect candidates for creating lossless transmit-arrays which are reflectionless at all frequencies and allow full control over transmission wavefronts.
Indeed, let us consider a lossless wideband reflectionless metasurface which can be realized using the same helical inclusions made of a low-loss material. Assuming the Lorentz polarizability model (\ref{eq:q6}) for both $\alpha_{\rm ee}$ and $\alpha_{\rm mm}$, the effective polarizabilities  of lossless balanced inclusions take the form 
\e {1\over \eta_0\widehat {\alpha}_{\rm ee}}={1\over \widehat {\alpha}_{\rm mm}/\eta_0}={\omega_{\rm e}^2-\omega^2\over A}-{i\omega \over 2S},\f
as follows from (\ref{eq:q3}) and (\ref{eq:q4}). The amplitude of the transmission coefficient is identically equal to unity, but its phase 
\e {\rm phase}\left(E_{\rm t}/E_{\rm inc}\right)=2\arctan \left({\omega \over 2S}{A\over {\omega_{\rm e}^2-\omega^2}}\right)\f
can be fully controlled by choosing the unit-cell area $S$ and the inclusion parameters $A$ and $\omega_{\rm e}$, still maintaining zero reflection property at all frequencies. Adjusting the level of dissipation loss in the inclusions, we can fully control also the amplitude of the transmitted wave. In contrast,  known realizations of Huygens' metasurfaces for controlling transmitted waves \cite{Huygens1,Huygens2,Huygens3} are perfectly matched only at one frequency. 

In this paper we have designed Huygens' metasheet absorbers which fully absorb incident radiation of one frequency, being totally transparent for radiation of other frequencies. This regime implies that the inclusions of the absorbers are equally strongly polarized electrically and magnetically, in as wide frequency range as possible. 
This is achieved by using one and the same resonant mode of a chiral resonator to create both electric and magnetic responses with identical spectral dispersions. From the theoretical point of view, the proposed absorber is probably the first microwave realization of a dispersive and lossy structure which does not reveal its dispersive and lossy nature when observed in reflections at any frequency. 
Although we have designed the prototypes operating in the microwave frequency range, our generic concept of Huygens' metasurface absorbers can be applied over the entire electromagnetic spectrum.

The unique property of zero reflectivity in an ultra-wide frequency range combined with frequency-selective absorption offers exciting possibilities in applications, allowing creation of perfect electrically thin band-stop filters for radiation of an arbitrary frequency. The off-band transparency of the absorbers allows one to construct various complex multi-frequency filters, combining in a parallel stack metasurfaces resonating at different frequencies. The neighbouring metasurfaces would not disturb the performance of one another, and overall thickness of such multi-layer structure would be still of the order of one wavelength or less. Another exciting possibility of implementation of the proposed absorbers lies in designing new types of ``invisible'' bolometers and sensors. Using the multi-layer Huygens' metasurface absorber, it becomes possible to design a single bolometer that measures power of incident radiation of {\it different} spectral lines {\it at the same time}. Moreover, the narrow-band response of the proposed absorbers makes them ideal candidates for implementation in bolometer arrays in astronomy at millimeter wavelengths.


\begin{acknowledgments}
This work was supported in part by the Nokia Foundation. The authors would like to thank Anu Lehtovuori and Mikko Honkala for helpful discussions of constant resistance circuits.
\end{acknowledgments}



\begin{thebibliography}{00}

\bibitem{perfect} 
N.~I.~Landy, S.~Sajuyigbe, J.~J.~Mock, D.~R.~Smith, and W.~J.~Padilla, {\it Perfect Metamaterial Absorber}, Phys. Rev. Lett. 100, 207402 (2008).

\bibitem{infrared} 
N.~Liu, M.~Mesch, T.~Weiss, M.~Hentschel, and H.~Giessen, {\it Infrared Perfect Absorber and Its Application As Plasmonic Sensor}, Nano Lett. 10, 2342 (2010).

\bibitem{soukoulis} 
M.~Diem, T.~Koschny, and C.~M.~Soukoulis, {\it Wide-angle Perfect Absorber/Thermal Emitter in the Terahertz Regime}, Phys. Rev. B 79, 033101 (2009).

\bibitem{munk2} 
B.~A.~Munk, {\it Jaumann and Circuit Analog Absorbers in Frequency-Selective Surfaces: Theory and Design} (Wiley, New York, 2000).

\bibitem{Rozanov} 
K.~N.~Rozanov, {\it Ultimate Thickness to Bandwidth Ratio of Radar
Absorbers}, IEEE Trans. Antennas Propag. 48, 1230 (2000).

\bibitem{bounds_tr} 
M.~Gustafsson, C.~Sohl, C.~Larsson, and D.~Sj\"oberg, {\it
Physical Bounds on the All-Spectrum Transmission Through Periodic Arrays}, 
Europhys. Lett. 87, 34002 (2009).

\bibitem{bounds_tr2} 
M.~Gustafsson, I.~Vakili, S.~Keskin, D.~Sj\"oberg, and 
C.~Larsson, {\it Optical Theorem and Forward Scattering Sum Rule
for Periodic Structures}, IEEE Trans. Antennas Propag. 60, 3818 (2012).

\bibitem{bounds_Fano} 
R.~M.~Fano, {\it Theoretical Limitations on the Broadband Matching
of Arbitrary Impedance}, Technical Report 41 (1948).  

\bibitem{kerker} 
M.~Kerker, D.-S.~Wang, and C.~L.~Giles, {\it Electromagnetic Scattering by Magnetic Spheres}, J. Opt. Soc. Am. 73, 765 (1983).

\bibitem{samofalov} 
I.~V.~Semchenko, S.~A.~Khakhomov, and A.~L.~Samofalov, {\it Helices of Optimal Shape for Nonreflecting Covering}, Eur. Phys. J. Appl. Phys. 49, 33002 (2010).

\bibitem{ozbay} 
E.~Saenz, I.~Semchenko, S.~Khakhomov, K.~Guven, R.~Gonzalo, E.~Ozbay, and S.~Tretyakov, {\it Modeling of Spirals with Equal Dielectric, Magnetic, and Chiral Susceptibilities}, Electromagnetics 28, 476 (2008).

\bibitem{sks} 
I.~V.~Semchenko, S.~A.~Khakhomov, and A.~L.~Samofalov, {\it Optimal Helix Shape: Equality of Dielectric, Magnetic, and Chiral Susceptibilities}, Russ. Phys. J 52, 472 (2009).

\bibitem{Huygens1} 
F.~Monticone, N.~M.~Estakhri, and A.~Al\`u, {\it Full Control of Nanoscale Optical Transmission with a Composite Metascreen}, Phys. Rev. Lett. 110, 203903 (2013).

\bibitem{Huygens2} 
C.~Pfeiffer and A.~Grbic, {\it Metamaterial Huygens' Surfaces: Tailoring Wave Fronts with Reflectionless Sheets}, Phys. Rev. Lett. 110, 197401 (2013).

\bibitem{Huygens3} 
M.~Decker, I.~Staude, M.~Falkner, J.~Dominguez, D.~N.~Neshev, I.~Brener, T.~Pertsch, and Y.~S.~Kivshar, {\it High-Efficiency Dielectric Huygens’ Surfaces}, Adv. Optical Mater., doi: 10.1002/adom.201400584 (2015).

\bibitem{ref1} 
C.~G.~Hu, X.~Li, Q.~Feng, X.~N.~Chen, and X.~G.~Luo, {\it Introducing Dipole-Like Resonance into Magnetic Resonance to Realize Simultaneous Drop in Transmission and Reflection at Terahertz Frequency}, J. Appl. Phys. 108, 053103 (2010).

\bibitem{ref2} 
H.~Tao, N.~Landy, C.~Bingham, X.~Zhang, R.~Averitt, and W.~Padilla, {\it A Metamaterial Absorber for the Terahertz Regime: Design, Fabrication and Characterization}, Opt. Express 16, 7181 (2008).

\bibitem{ref3} 
G.~R.~Keiser, A.~C.~Strikwerda, K.~Fan, V.~Young, X.~Zhang, and R.~D.~Averitt, {\it Decoupling Crossover in Asymmetric Broadside Coupled Split-Ring Resonators at Terahertz Frequencies}, Phys. Rev. B 88, 024101 (2013).

\bibitem{ref4} 
Y.~Cheng, H.~Yang, Z.~Cheng, and B.~Xiao, {\it A Planar Polarization-Insensitive Metamaterial Absorber}, Photon. Nanostr. Fundam. Appl. 9, 8 (2011).

\bibitem{7padilla} 
N.~I.~Landy, C.~M.~Bingham, T.~Tyler, N.~Jokerst, D.~R.~Smith, and W.~J.~Padilla, {\it Design, Theory, and Measurement of a Polarization-Insensitive Absorber for Terahertz Imaging}, Phys. Rev. B 79, 125104 (2009).

\bibitem{bilotti} 
F.~Bilotti, A.~Toscano, K.~B.~Alici, E.~Ozbay, and L.~Vegni, {\it Design of Miniaturized Narrowband Absorbers Based on Resonant-Magnetic Inclusions}, IEEE Trans. Electromagn. Compat. 53, 63 (2011).

\bibitem{absorption} 
Y.~Ra'di, V.~S.~Asadchy, and S.~A.~Tretyakov, {\it Total Absorption of Electromagnetic Waves in Ultimately Thin Layers}, IEEE Trans. Antennas Propag. 61, 4606 (2013).

\bibitem{serdukov} 
A.~N.~Serdyukov, I.~V.~Semchenko, S.~A.~Tretyakov, and A.~Sihvola, {\it Electromagnetics of Bi-anisotropic Materials: Theory and Applications} (Gordon and Breach Science, Amsterdam, 2001).

\bibitem{CP} 
M.~Li, L.~Guo, J.~Dong, and H.~Yang, {\it An Ultra-thin Chiral Metamaterial Absorber
with High Selectivity for LCP and RCP Waves}, J. Phys. D: Appl. Phys. 47, 185102 (2014).

\bibitem{Teemu} 
T.~Niemi, A.~O.~Karilainen, and S.~A.~Tretyakov, {\it Synthesis of Polarization
Transformers}, IEEE Trans. Antennas Propag. 61, 3102 (2013).

\bibitem{tretyakov} 
S.~A.~Tretyakov, {\it Analytical Modeling in Applied Electromagnetics} (Artech House, Norwood, 2003).

\bibitem{core-shell1} 
R.~Paniagua-Dominguez, F.~Lopez-Tejeira, R.~Marques, and J.~A.~Sanchez-Gil, {\it Metallo-dielectric Core–shell Nanospheres as Building Blocks for Optical Three-dimensional Isotropic Negative-index Metamaterials}, New J. Phys. 13, 123017 (2011).

\bibitem{core-shell2}  
D.~Morits and C.~R.~Simovski, {\it Isotropic Negative Refractive Index at Near Infrared}, J. Optics 14, 125102 (2012).

\bibitem{arxiv} 
S.~Campione, L.~I.~Basilio, L.~K.~Warne, and M.~B.~Sinclair, {\it Tailoring Dielectric Resonator Geometries for Directional Scattering and Huygens’ Metasurfaces}, Opt. Express 23, 2293 (2015).

\bibitem{Padilla} 
C.~M.~Watts, X.~Liu, and W.~J.~Padilla, {\it Metamaterial Electromagnetic Wave Absorbers}, Adv. Mater. 24, OP98 (2012).

\bibitem{antenna_model}
S.~A. Tretyakov, F. Mariotte, C.~R. Simovski, T.~G. Kharina, and J.-P. Heliot, {\it Analytical Antenna Model for Chiral Scatterers: Comparison with Numerical and Experimental Data}, IEEE Trans. Antennas Propag. 44, 1006 (1996).

\bibitem{chiral} 
B.~Wang, J.~Zhou, T.~Koschny, M.~Kafesaki, and C.~M.~Soukoulis, {\it Chiral Metamaterials: Simulations and Experiments}, J. Opt. A: Pure Appl. Opt. 11, 114003 (2009).

\bibitem{alu} 
Y.~Zhao, M.~A.~Belkin, and A.~Al\`u, {\it Twisted Optical Metamaterials for Planarized Ultrathin Broadband Circular Polarizers}, Nat. Commun. 3, 870 (2012).

\bibitem{conference2} 
I.~A.~Faniayeu, V.~S.~Asadchy, T.~A.~Dziarzhauskaya, I.~V.~Semchenko, and S.~A.~Khakhomov, {\it in Proceedings of the 8th International Congress on
Advanced Electromagnetic Materials in Microwaves and Optics, 2014}, p.~112.

\bibitem{fabrication1} 
J.~K.~Gansel, M.~Thiel, M.~S.~Rill, M.~Decker, K.~Bade, V.~Saile, G.~Freymann, S.~Linden, and M.~Wegener, {\it Gold Helix Photonic Metamaterial as Broadband Circular Polarizer}, Science 325, 1513 (2009).

\bibitem{fabrication2} 
A.~Radke, T.~Gissibl, T.~Klotzb\"ucher, P.~V.~Braun, and H.~Giessen, {\it Three-Dimensional Bichiral Plasmonic Crystals Fabricated by Direct Laser Writing and Electroless Silver Plating}, Adv. Mater. 23, 3018 (2011).

\bibitem{samofalov2} 
I.~V.~Semchenko, S.~A.~Khakhomov, and A.~L.~Samofalov, {\it Transformation of the Polarization of Electromagnetic Waves by Helical Radiators}, J. Commun. Technol. El. 52, 850 (2007).

\bibitem{conference1} 
V.~S.~Asadchy, I.~A.~Faniayeu, Y.~Ra’di, I.~V.~Semchenko, and S.~A.~Khakhomov, {\it in Proceedings of the 7th International Congress on
Advanced Electromagnetic Materials in Microwaves and Optics, 2013}, p.~244.

\bibitem{polarizability} 
V.~S.~Asadchy, I.~A.~Faniayeu, Y.~Ra’di, and S.~A.~Tretyakov, {\it Determining Polarizability Tensors for an Arbitrary Small Electromagnetic Scatterer}, Phot. Nano. Fund. Appl. 12, 298 (2014).

\bibitem{ANSYS} 
ANSYS HFSS, 2014: www.ansoft.com.


\bibitem{zobel}
O.~J.~Zobel, {\it Distortion  Correction  in Electrical  Circuits with Constant
Resistance Recurrent Networks}, Bell Syst. Tech. J. 7, 348 (1928).

\bibitem{norton}
E.~L.~Norton, {\it Constant Resistance Networks with Applications to Filter
Groups}, Bell Syst. Tech. J. 16, 178 (1937).

\bibitem{method} 
D.~K.~Ghodgaonkar, V.~V.~Varadan, and V.~K.~Varadan, {\it A Free-Space Method for Measurement of Dielectric Constants and Loss Tangents at Microwave Frequencies}, IEEE Trans. Instrum. Meas. 37, 789 (1989).

































%
%
%










\end{thebibliography}
\end{document}